%% file: main.tex
\documentclass[sigconf]{acmart}

\AtBeginDocument{%
  \providecommand\BibTeX{{%
    \normalfont B\kern-0.5em{\scshape i\kern-0.25em b}\kern-0.8em\TeX}}}

\setcopyright{acmcopyright}
\copyrightyear{2018}
\acmYear{2018}
\acmDOI{XXXXXXX.XXXXXXX}

\acmConference[Conference acronym 'XX]{Make sure to enter the correct
  conference title from your rights confirmation emai}{June 03--05,
  2018}{Woodstock, NY}

\acmBooktitle{Woodstock '18: ACM Symposium on Neural Gaze Detection,
 June 03--05, 2018, Woodstock, NY} 
\acmPrice{15.00}
\acmISBN{978-1-4503-XXXX-X/18/06}

\input{macro.tex}

\begin{document}

\title{GazeNoter: Co-Piloted AR Note-Taking via Gaze Selection of LLM Suggestions to Match Users’ Intentions} 

\author{Hsin-Ruey Tsai}
\authornote{Corresponding author}
\affiliation{%
  \institution{National Chengchi University}
  \city{Taipei}
  \country{Taiwan}}
\email{hsnuhrt@gmail.com}

\author{Shih-Kang Chiu}
\affiliation{%
  \institution{National Chengchi University}
  \city{Taipei}
  \country{Taiwan}}
\email{di9italmaximalism@gmail.com}

\author{Bryan Wang}
\affiliation{%
  \institution{Adobe Research}
  \city{Seattle}
  \state{WA}
  \country{United States}}
  \affiliation{
  \institution{University of Toronto}
  \city{Toronto}
  \country{Canada}}
\email{bryanw@adobe.com}

\renewcommand{\shortauthors}{Tsai et al.}

\begin{abstract}
  Note-taking is critical during speeches and discussions, serving not only for later summarization and organization but also for real-time question and opinion reminding in question-and-answer sessions or timely contributions in discussions.
  Manually typing on smartphones for note-taking could be distracting and increase cognitive load for users.
  While large language models (LLMs) are used to automatically generate summaries and highlights, the content generated by artificial intelligence (AI) may not match users' intentions without user input or interaction.
  Therefore, we propose an AI-copiloted augmented reality (AR) system, GazeNoter, to allow users to swiftly select diverse LLM-generated suggestions via gaze on an AR headset for real-time note-taking.
  GazeNoter leverages an AR headset as a medium for users to swiftly adjust the LLM output to match their intentions, forming a user-in-the-loop AI system for both within-context and beyond-context notes.
  We conducted two user studies to verify the usability of GazeNoter in attending speeches in a static sitting condition and walking meetings and discussions in a mobile walking condition, respectively.
  
\end{abstract}

\ccsdesc[500]{Human-centered computing~Mixed / augmented reality}
\ccsdesc[300]{Human-centered computing~Natural language interfaces}
\ccsdesc[300]{Human-centered computing~Text input}

\keywords{note-taking, augmented reality, large language models, artificial intelligence, gaze input, wearable devices}

\begin{teaserfigure}
\begin{center}
  \includegraphics[width=1\textwidth]{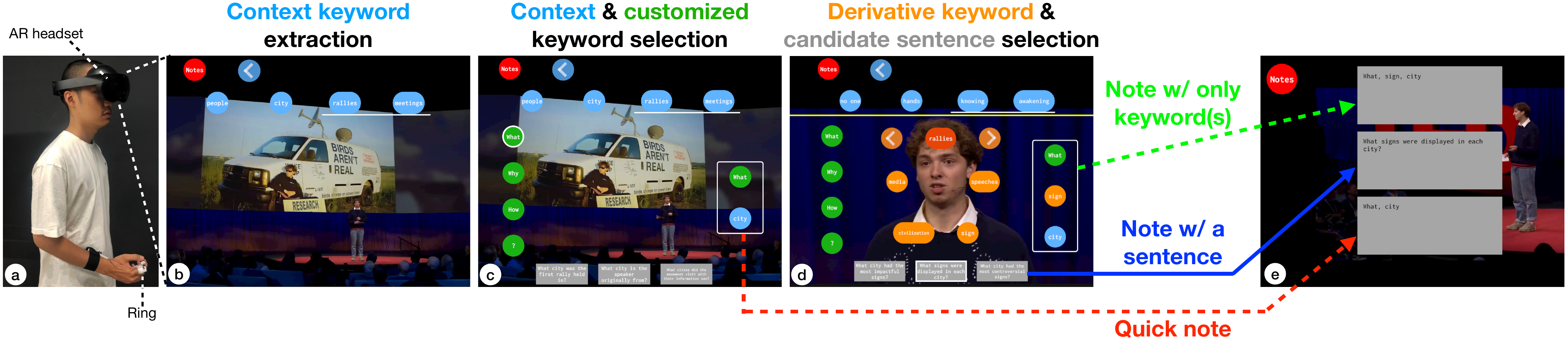}
  \caption{Users swiftly select LLM-generated suggestions via gaze on an AR headset with a ring for real-time note-taking in GazeNoter. 
  (a) An AR headset and a ring are worn for note-taking in a speech. 
  (b) Extracted context keywords from the latest sentence of the speech. 
  (c) Users could select context and customized keywords, and candidate sentences are then automatically generated. 
  (d) Users could explore and select derivative keywords beyond the context of the speech and select a candidate sentence best matching the intentions to record as a note. 
  (e) Users could review the recorded notes, which may be from three processes.
  (Blue) Normally, a sentence is recorded as a note. (Green) If no candidate sentences match users' intentions, users could record all selected keywords as a note. (Red) If users need to take a note hastily, users could select context (or also customized) keywords, skipping step (d), to record these as a \textit{quick note}.}
  \Description{Users swiftly select LLM-generated suggestions via gaze on an AR headset with a ring for real-time note-taking in GazeNoter. (a) An AR headset and a ring are worn for note-taking in a speech. (b) Extracted context keywords from the latest sentence of the speech. (c) Users could select context and customized keywords, and candidate sentences are then automatically generated. (d) Users could explore and select derivative keywords beyond the context of the speech and select a candidate sentence best matching the intentions to record as a note. (e) Users could review the recorded notes, which may be from three processes. (Blue) Normally a sentence is recorded as a note. (Green) If no candidate sentences match users’ intentions, users could record all selected keywords as a note. (Red) If users need to take a note hastily, users could select context (or also customized) keywords, skipping step (d), to record these as a quick note.}
  \label{fig:teaser}
  \end{center}
\end{teaserfigure}

\received{20 February 2007}
\received[revised]{12 March 2009}
\received[accepted]{5 June 2009}

\maketitle

\section{Introduction}

\begin{figure}
\begin{center} 
\includegraphics[width=0.8\linewidth]{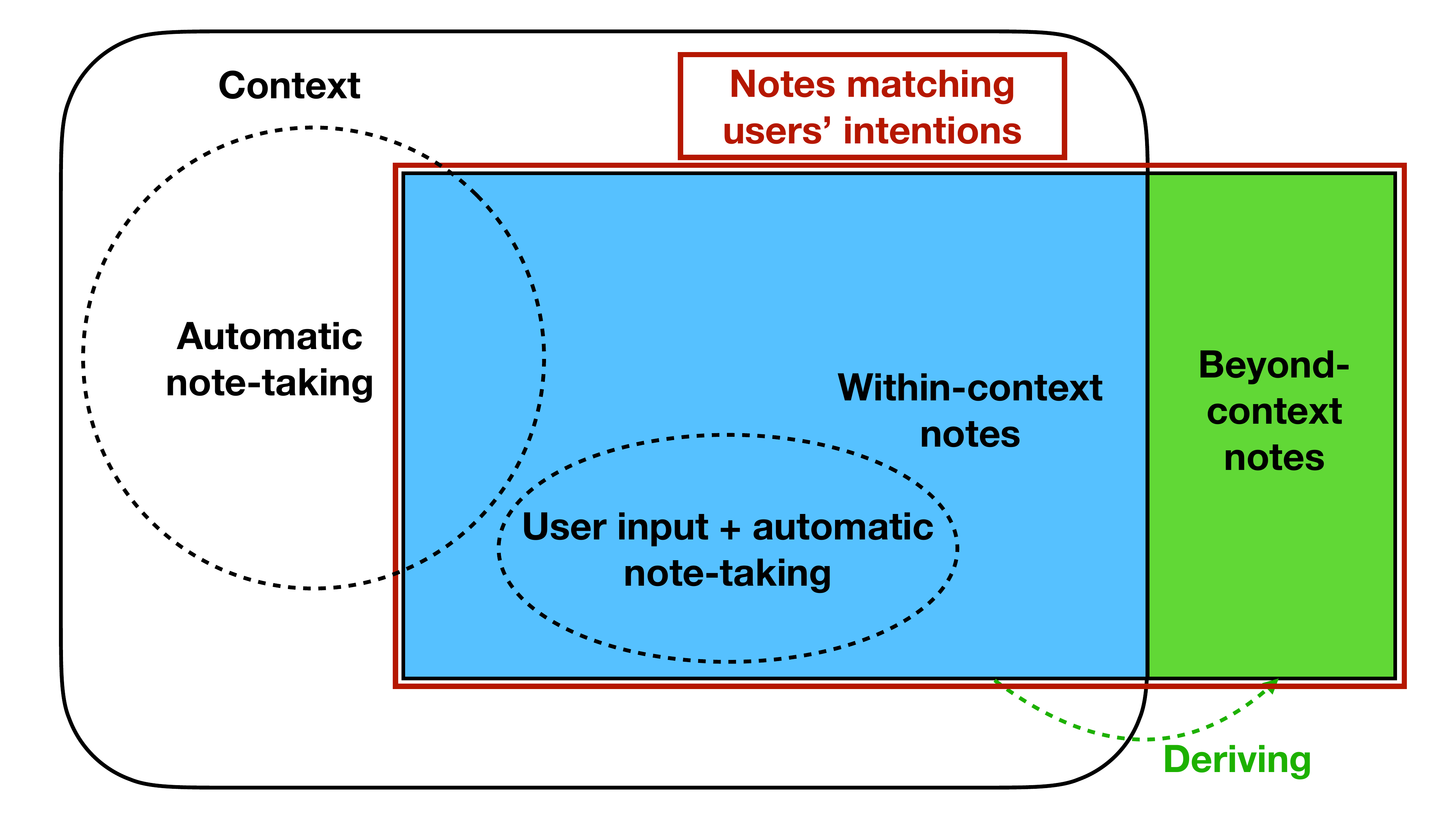}
\caption{
The blue part represents the notes users want to directly record as hearing the context, which is defined as within-context notes. The green part represents notes that combine users' insights, defined as beyond-context notes. Achieving both within-context and beyond-context notes ensures that the notes align with users' intentions.
}
\Description{The blue part represents the notes users want to directly record as hearing the context, which is defined as within-context notes. The green part represents notes that combine users' insights, defined as beyond-context notes. Achieving both within-context and beyond-context notes ensures that the notes align with users' intentions.}
\label{fig:beyond}
\vspace*{-6pt}
\end{center}
\end{figure}

Note-taking is crucial during speech-based activities, such as in speeches and discussions.
Besides summarizing content or organizing thoughts, it can act as real-time reminders for questions or opinions,  particularly during question-and-answer (Q\&A) sessions or when contributing timely to conversations. 
Common methods for note-taking include longhand notes and smartphone text input. 
However, manual note-taking can divert attention from the primary activity, distracting users and increasing their cognitive load~\cite{piolat2005cognitive, mosleh2016challenges}. This issue becomes more pronounced in mobile scenarios, such as taking notes during walking meetings. 
To address these challenges, prior research has utilized natural language processing (NLP) and large language models (LLMs) to automatically generate summaries~\cite{xu2022semantic, shi2018meetingvis}, and highlights~\cite{roy2021note, laban2023designingto} from a transcript context for note-taking. 
However, without user input, the auto-generated notes may not always align with the users' intentions.
To tackle this issue, recent studies~\cite{kalnikaite2012markup, haliburton2023walking} have sought to involve user input in the automatic note-taking process by enabling users to highlight crucial statements during meetings. 
Nevertheless, these methods still fall short as they generate notes only within the context of the transcript but cannot produce notes or inferences that go beyond the context, which typically requires insights or inputs specific to the users, as illustrated in \figname~\ref{fig:beyond}.
The importance of this type of derivative notes, which we refer to as \textit{beyond-context notes}, has been emphasized by previous work in note-taking research~\cite{fang2022understanding}.

A nuanced system that incorporates user input into automatic note-taking, ensuring the alignment of generated content with users' intentions, and providing support for both within-context and beyond-context notes is then required (the red block in \figname~\ref{fig:beyond}).
Artificial intelligence (AI) and LLMs excel in diverse suggestion generation, and user input achieves more precise users' intentions. 
By integrating user input into AI, forming a user-in-the-loop interaction paradigm, users can copilot with AI to adjust and generate more precise and desirable outcomes, which is leveraged in some works~\cite{wang2022record} but not for note-taking. 
Implementing this user-in-the-loop AI system for real-time note-taking requires a medium that facilitates swift, subtle, and low-distraction input and output manners for displaying and adjusting LLM-generated content to reduce users' cognitive load and match users’ intentions. 
Augmented reality (AR) headsets with see-through displays, offering gaze selection, are ideal for this purpose, reducing the need to look down at smartphones and minimizing distraction~\cite{cai2023paraglassmenu, janaka2022paracentral}. 
Furthermore, AR interactions, enhanced by AI, can display content around real objects or humans, ensuring convenience and low distraction~\cite{liu2023visual}. 
Hence, an AR headset is proposed as the medium for this AI system, considering the expected rise in AR device usage in the future.

We present an AI-copiloted AR system, GazeNoter, enabling users to swiftly select LLM-generated suggestions via gaze on an AR headset for real-time note-taking. 
GazeNoter extracts keywords from the context of real-time audio transcripts via the LLM.
Users can select these context keywords, or prompt the LLM to derive more keywords from the selected context keywords for further selection, aligning closely with users' intentions.
Using the selected keywords, the LLM then organizes candidate sentences that might encapsulate the desired notes for users to record the notes. 
By leveraging the three capabilities of the LLM, including extraction, derivation and organization, and efficient gaze selection for keyword and sentence selection, this AI-copilot AR system achieves real-time note-taking with low distraction and cognitive load, matching users' intentions for both within-context and beyond-context notes. 
While individual components, including LLMs for note-taking and AR gaze interactions, have been explored before, GazeNoter uniquely integrates these for mobile, real-time note-taking without voice or typing input.
We conducted two user studies to compare the performance of GazeNoter with other note-taking methods, respectively verifying its usability in two scenarios: (1) attending speeches in a static sitting condition, similar to lecture scenarios widely studied in prior research~\cite{mosleh2016challenges, piolat2005cognitive} but with more time constraints for note-taking; and (2) walking meetings and discussions in a mobile walking condition, an emerging use case~\cite{damen2020hub,damen2020understanding, haliburton2023walking} with additional challenges due to mental pressure and the walking environment.

Taken together, our work makes the following contributions:
\begin{enumerate}
    \item The design and implementation of GazeNoter, a real-time, user-in-the-loop note-taking system integrating the LLM and AR, enabling swift and low-distraction note generation for both within-context and beyond-context notes while matching users' intentions.

    \item The results from two user studies, demonstrated GazeNoter's effectiveness in static and mobile scenarios over baseline comparisons, respectively. 
    This effectiveness included reduced distraction and cognitive load, enhanced subtlety and usability, and improved resulting notes that better match users' intentions and more effectively remind users. 
    In the mobile scenario, GazeNoter further improved frustration management, physical effort, and social acceptance.

\end{enumerate}

\section{Related Work}

Our system design involves augmenting AR with natural language, note-taking approaches, user-in-the-loop NLP systems and gaze selection on headsets. Therefore, we discuss previous research about these in this section.
\subsection{Augmenting AR with Natural Language}
Utilizing NLP to interpret textual data from speech-to-text or camera-based text recognition paves the way for innovative context-aware interactions in AR.
RealityTalk~\cite{liao2022realitytalk} enables users to interact with virtual elements through speech in augmented presentations. 
SpeechBubbles~\cite{peng2018speechbubbles} uses real-time speech transcription as textual aids to enhance captioning experiences in AR for hard-of-hearing people in conversations.
Moreover, LLMs are employed to further improve comprehension and generation of natural language, enabling more nuanced and responsive systems.
Visual Captions~\cite{liu2023visual} leverages a fine-tuned LLM to display relevant visual content during open-vocabulary conversations.
ARFit~\cite{mandic2023arfit} offers an AR system that combines pose tracking and an LLM to deliver feedback on exercise movements, emulating expert advice for maintaining correct posture.
VisionARy~\cite{lee2023VisionARy} proposes a real-time English speaking practice system that integrates an LLM and object recognition with an AR headset to provide an interactive and contextualized English-speaking experience in everyday life.
However, leveraging AI and LLMs to achieve note-taking in AR, especially in real time, is still underexplored but presents promising for further exploration. 

\subsection{Note-Taking Approaches}

A variety of research has focused on facilitating the process of note-taking in lectures ~\cite{kam2005livenotes, silvestre2014tsaap, plaue2012group, kaminski2016learning}, meetings ~\cite{damen2020hub, banerjee2006smartnotes, banerjee2007segmenting, exposito2017unobtrusive}, online learning videos~\cite{nguyen2016gaze, liu2019notestruct, cao2022videosticker, samson2018lessonware} and other textual contents~\cite{laban2023designing, rajaram2022paper}. 
Assistive features are provided to enable more sophisticated interactions in note-taking systems.
NoteLink~\cite{srinivasa2021notelink} retrieves relevant lecture videos based on students' handwritten notes.
NoteCoStruct~\cite{fang2021notecostruct} fosters a sense of learning community through note-sharing from previous learners.
CoNotate~\cite{palani2021conotate} provides query suggestions based on the notes to assist exploratory searches in unfamiliar domains.
However, text entry or handwriting is required in these works, which could divert users' attention from ongoing activities, increasing cognitive load~\cite{piolat2005cognitive, mosleh2016challenges}.
Besides, although voice recording is also used to take audio notes~\cite{khan2019gaze,khan2022type}, which is proven more effective than typing notes using keyboard~\cite{khan2022type}, broadcasting to listen to the notes is inappropriate in activities like speeches and discussions.

On the other hand, several works focus on automatically generating summaries~\cite{xu2022semantic, yang2022catchlive, li2023improving} and highlights~\cite{laban2023designingto, hautasaari2014catching} to provide meaningful contents, especially with the development of LLMs. 
TalkTraces ~\cite{Chandrasegaran2019talktraces}and MeetingVis~\cite{shi2018meetingvis} focus on real-time visualized summarization tools designed to facilitate productive group discussions by recognizing topics and providing a visualized overview of the agenda in real-time.
Li~\etal~\cite{li2021hierarchical} propose a system to generate different levels of summary and allow users to browse and navigate the content more efficiently. 
Beyond Text Generation~\cite{dang2022beyond} generates summaries to provide writers with an overview of their writing from an external perspective during the writing process. 
These automatic note-taking methods are effective in producing summarized notes directly from the context of the transcripts, reducing cognitive load, but without users' participation in the note-taking process, the notes within the context might not always align with those the users intend to record.
Furthermore, they cannot generate beyond-context notes derived by users from the context, a factor whose importance has been shown in~\cite{fang2022understanding}. 
On the other hand, The Walking Talking Stick~\cite{haliburton2023walking} and Markup as you talk~\cite{kalnikaite2012markup} enable users to press a physical button to highlight critical statements during a meeting to discern and document users' intentions.
However, they might capture only segments of the transcript for within-context notes but not for beyond-context notes.

\subsection{User-in-the-Loop NLP Systems}

Instead of only using training models, previous research integrates an interactive process into AI systems and allows users to adjust and fine-tune the result with immediate feedback~\cite{jin2017elasticplay, zhou2022gesture, wu2022ai, lu2023readingquizmaker, brade2023promptify}, which is also called the user-in-the-loop interaction paradigm. 
Such an interactive design enables the systems to generate more tailored, user-centered results. 
The integration of user-in-the-loop interaction paradigm and with NLP has been effectively utilized across a variety of applications.
ROPE~\cite{wang2022record} proposes an automatic audio shortening system based on the semantics and duration of the audio clip and can optimize results after users specify sentences to include or exclude.
Crosspower~\cite{xia2020crosspower} explores the utility of transforming linguistic structures extracted from written content into graphic content and enables users to interact with language structures and their graphic correspondences to create graphic effects.
ConceptEVA~\cite{zhang2023concepteva} proposes a document summarization system to achieve the customization of long document summaries through an interactive visual analysis and NLP techniques.

Recent research has further incorporated user-in-the-loop with LLMs to improve the comprehension and generation capabilities of natural language in AI systems.
Several research~\cite{mirowski2023co, chung2022talebrush} offer users the flexibility to interact with the script generation process by LLMs and enrich the co-creation experience with LLMs in writing.
Graphologue~\cite{jiang2023graphologue} and Sensecape~\cite{suh2023sensecape} propose systems that enable users to interact with LLMs through non-linear, node-link dialogues and allow users to dynamically tailor the graphical representation of information, offering flexibility in exploring and understanding complex knowledge. 
Promptify~\cite{brade2023promptify} introduces an interactive tool that enhances the prompt creation process for text-to-image models by offering a suggestion engine and a flexible interface for easy exploration and refinement.
By integrating user input and AI, user-in-the-loop AI systems are able to generate more precise outcomes matching users' intentions.
Although these works do not focus on note-taking, we aim to leverage the merits to enable a novel real-time AR note-taking system with low cognitive load, distraction, and matching users' intentions for both within-context and beyond-context notes.

\subsection{Gaze Selection and Layout on Headsets}

Gaze is a common input method on VR/AR headsets.
Rapid eye movement enables swift gaze selection, but performing confirmation using gaze is challenging. 
Previous research proposes several gaze-only target selection approaches, including dwelling~\cite{pi2017probabilistic,thaler2013best}, gaze gestures~\cite{dybdal2012gaze, sidenmark2020outline}, and eye vergence~\cite{kudo2013input,kirst2016verge,ahn2020verge}.
On the other hand, others combine gaze and auxiliary modalities for gaze selection, including head movement~\cite{sidenmark2019eye,sidenmark2021radi,wei2023predicting}, eyelid movement~\cite{yi2022deep}, tongue gesture~\cite{gemicioglu2023gaze} and hand gesture~\cite{pfeuffer2017gaze+} on headsets. 
Furthermore, some research focuses on designing a gaze-based menu to achieve more stable selections and reduce eye fatigue~\cite{ahn2021stickypie, kim2022lattice, choi2022kuiper}. 
In gaze selection, interacting with virtual elements via gaze on AR headsets without interrupting conversations is another critical issue.
StARe~\cite{rivu2020stare} and {Glanceable AR~\cite{lu2020glanceable} focus on designing progressively revealing information on demand to minimize distraction using gaze on headsets. 
ParaGlassMenu~\cite{cai2023paraglassmenu} and Paracentral and Near-Peripheral Visualizations~\cite{janaka2022paracentral} concentrated on designing the layout on OHMDs. 
Their goal was to maintain undiverted attention while performing other tasks, such as controlling digital devices or displaying information, on the OHMD.
Based on these prior works, GazeNoter leverages the swift gaze selection in AR note-taking with a proper layout for social acceptance and subtlety.

\section{GazeNoter}

We propose GazeNoter for real-time note-taking with low distraction and cognitive load, and matching users' intentions in speech-based activities for within-context and beyond-context notes.
In terms of inferring users' intentions, brain-computer interfaces (BCI) could be the most desirable approach.
However, current BCI techniques still cannot achieve precise brain control or mind reading as in science-fiction novels and movies.
GazeNoter leverages AI as a tool for human-like thinking and utilizes an AR headset as a medium for users to swiftly adjust the generated output, and this user-in-the-loop AI system bypasses the BCI limits to achieve real-time note-taking matching users' intentions.

\subsection{Design Considerations}

To accomplish our goals, the following design considerations should be taken into account.

\begin{itemize}
    \item \textit{DC1. Real-Time System.}
    In speech-based activities, \eg speeches or discussions, taking notes for critical information, the essence of discourses, questions or opinions should be in a short period for the upcoming interactive Q\&A sessions or discussions.
    Therefore, a real-time note-taking system that enables users to quickly take notes and preserve their thoughts at the moment becomes crucial.
    This means that instead of typing or speaking complete sentences for taking notes, swift input and output manners, such as only a few selection steps, for the note-taking system are required.

    \item \textit{DC2. Matching Users' Intentions.}
    Without handwriting or text entry input, it is challenging to obtain users' intentions in note-taking.
    Although LLMs excel in summarization and highlighting for note-taking, the automatically generated content as within-context notes could be diverse and might not always match users' intentions since user input is not in the process.
    This also means that these automatic note-taking methods cannot generate beyond-context notes derived by users from the context. 
    Moreover, previous studies have underscored the critical role of notes that incorporate derivative words and go beyond the given context to improve the quality of notes. 
    Such notes are characterized as being constructive and interactive within the adaptive Interactive, Constructive, Active, and Passive (ICAP) framework~\cite{fang2022understanding}, which categorizes note quality to enhance the learning experience.
    By integrating user input into AI and forming a user-in-the-loop AI system, users can adjust the diverse LLM-generated output to converge closely to their intentions, generating more precise notes matching users' intentions. 
    For such a system, there is a trade-off between generating more precise notes matching users' intentions and avoiding complicated user input for swift input and lower distraction and cognitive load.
    
    \item \textit{DC3. Distraction and Cognitive Load.}
    Since users have to stay engaged in the ongoing activity while taking notes, note-taking via handwriting or smartphone text entry input could distract them and increase their cognitive load.
    The increased cognitive load may result in individuals taking incomplete notes~\cite{kiewra1989review}.
    Thus, note-taking methods with input and output manners that are either eyes-free or allow users to keep their visual attention on or around the ongoing activity are required.
    Eyes-free output methods, such as audio output, might be slow for real-time systems and have privacy issues.
    On the other hand, an always-worn AR headset with a see-through display to rapidly switch between virtual and real worlds for output that does not require users to look down at smartphones, allowing users to keep their visual attention on or around the ongoing activity could be an adequate medium.
    However, user input steps and time through the medium, and displayed content and layout on the medium might also affect users, potentially leading to distraction and cognitive load, which should be considered carefully.

    \item \textit{DC4. Subtlety.} 
    Several speech-based activities, requiring note-taking, demand the note-takers' participation and concentration.
    Body language or interactions like nodding or eye contact could be important, especially in discussions and conversations or even in speeches and lectures. 
    Using a smartphone could be indiscreet or impolite in such occasions.
    Therefore, the medium and its input and output manners for the user-in-the-loop note-taking system should be subtle.
    Although current AR headsets are still bulky, we envision that AR headsets could be reduced to the size of normal glasses like Google Glass.
    The see-through displays and gaze or microgestures on AR headsets could be considered as subtle input and output manners.
    
    \item \textit{DC5. Mobility.}
    Note-taking is required in static scenarios, such as seating or even with a desk, and in mobile scenarios, such as attending walking meetings~\cite{haliburton2023walking, damen2020hub, damen2020understanding}.
    Thus, a portable design, including lightweight devices and even hands-free interactions for mobility should be considered.
    
\end{itemize}

\begin{figure*}[ht]
\begin{center} 
\includegraphics[width=1\linewidth]{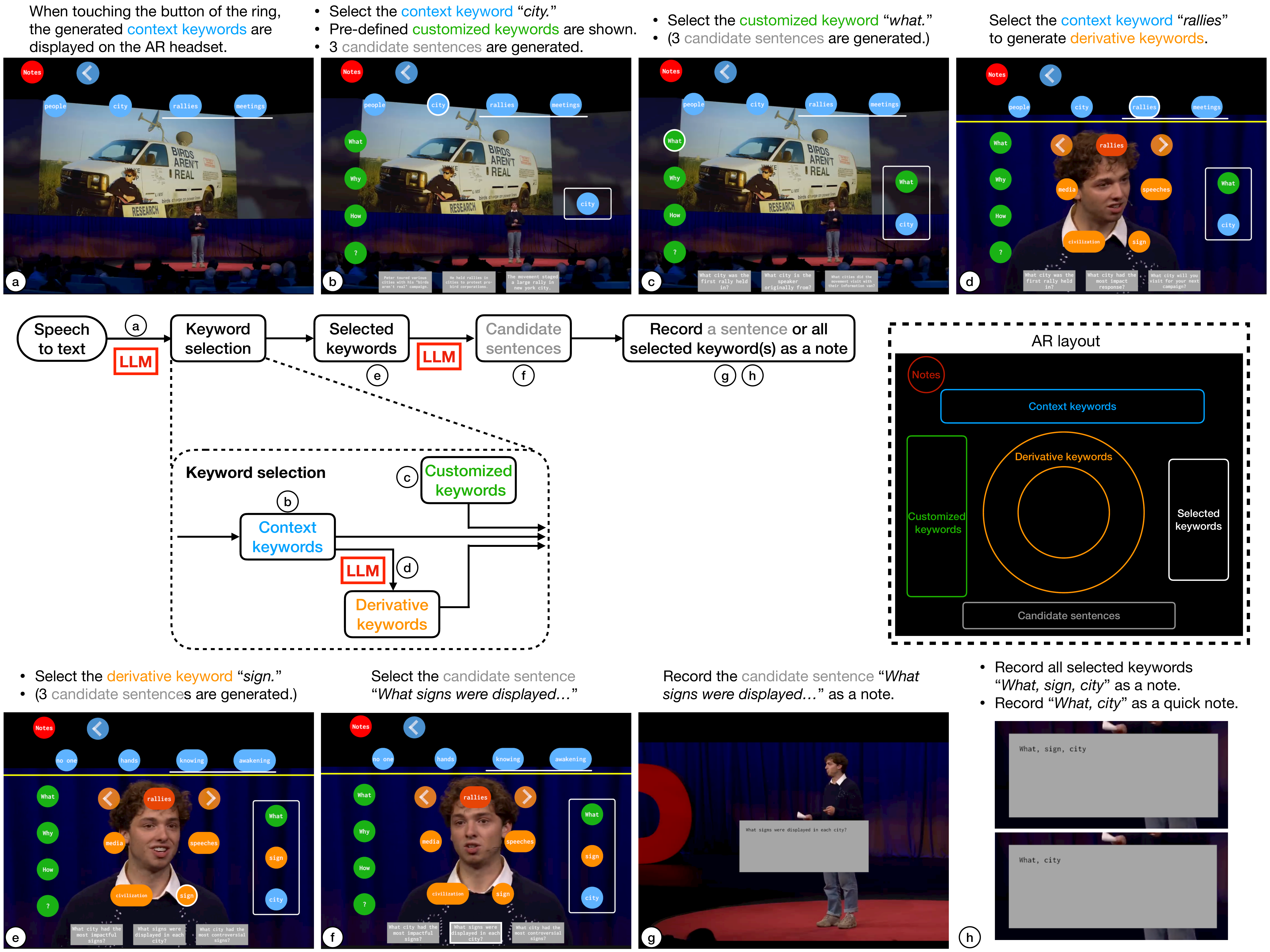}
\caption{(Middle) The flowchart of GazeNoter. (Right) The AR layout, displayed only when the ring is touched. (a) GazeNoter extracts context keywords from the latest sentence of the speech. 
(b) Once the user selects a context keyword, ``\textit{city}'', the pre-defined customized keywords are shown, and 3 candidate sentences are generated based on the context keyword. 
(c) The user selects a customized keyword, ``\textit{what}'', and the candidate sentences are updated accordingly. 
(d) If the user wants to take a beyond-context note and no desired keyword is among the context keywords, the user selects the most relevant context keyword, ``\textit{rallies}'', to generate derivative keywords. 
(e) The user selects a derivative keyword, ``\textit{sign}'', and the candidate sentences beyond the context are updated accordingly. 
(f) The user selects the candidate sentence best matching the intention, ``\textit{What signs were displayed...}'', to record as a note. 
(g) The recorded note is shown. 
(h) If no candidate sentences match the intention in step (f), the user could also record all selected keywords as a note (upper). 
If the user needs to take a note hastily, the user could select only context (or also customized) keywords to record these as a \textit{quick note} (lower), only from steps (a)(b) or (a)(b)(c).} 
\Description{(Middle) The flowchart of GazeNoter. (Right) The AR layout, which is shown only when the ring is touched. (a) GazeNoter extracts context keywords from the latest sentence of the speech. (b) Once the user selects a context keyword, ''city'', the pre-defined customized keywords are shown, and 3 candidate sentences are generated based on the context keyword. (c) The user selects a customized keyword, ''what'', and the candidate sentences are updated accordingly. (d) If the user wants to take a beyond-context note and no desired keyword is among the context keywords, the user selects the most relevant context keyword, ''rallies'', to generate derivative keywords. (e) The user selects a derivative keyword, ''sign'', and the candidate sentences beyond the context are updated accordingly. (f) The user selects the candidate sentence best matching the intention, ''What signs were displayed in each city?'', to record as a note. (g) The recorded note is shown. (h) If no candidate sentences match the intention in step (f), the user could also record all selected keywords as a note (upper). If the user needs to take a note hastily, users could select only context (or also customized) keywords to record these as a quick note (lower), only from steps (a)(b) or (a)(b)(c).}
\label{fig:flowchart}
\vspace*{-6pt}
\end{center}
\end{figure*}

\subsection{GazeNoter Features and Flow}
Based on the real-time system (DC1) and matching users' intentions (DC2) design considerations, a user-in-the-loop AI system with a few user selection steps is the desired interaction for real-time note-taking.
When users need to take quick notes, they usually record keywords instead of typing or speaking a complete sentence, as mentioned in~\cite{chen2021exploring}.
Furthermore, previous research has shown that keywords enable users to quickly understand the current context~\cite{son2023okay, yang2022catchlive, dang2022beyond}.
Therefore, the habit and merits of using keyword-based interaction in note-taking are leveraged in the GazeNoter design.
GazeNoter showcases the LLM-generated suggestions on the AR headset, enabling users to select and adjust these contents via gaze and ring input to take their desired notes. 
The AR gaze selection and system implementation details and design considerations are described in the following subsections.
Three capabilities of LLMs, including extraction, derivation, and organization, are used in this user-in-the-loop LLM system.
In GazeNoter, users select LLM-generated keywords and then select an LLM-generated sentence to record as a note, as shown in~\figname~\ref{fig:flowchart}.
In this keyword and sentence selection procedure, four essential features are leveraged, including the selection of three types of keywords, \textit{context keyword}, \textit{customized keyword}, and \textit{derivative keyword}, along with the selection of \textit{candidate sentence}.
Furthermore, one additional feature, \textit{note refinement}, as well as these four features constitute GazeNoter.
The four features are positioned close to the edges of the AR display or arranged in a circular layout, leaving the center blank for real-world activities (\figname~\ref{fig:flowchart} (right)).
All AR contents are displayed on the AR headset only when the button of the ring is touched for note-taking.
Users utilize gaze to select an item and click the button to confirm the selected item. 
Double-click is used as an alternative selection method.

\subsubsection{Context Keyword Selection}

The system captures a voice segment, subsequently displaying context keywords via the LLM keyword extraction in real-time from the latest sentence. 
As shown in~\figname~\ref{fig:flowchart} (a), these context keywords appear on the top section of the AR display in a 1 $\times$ 4 layout, at most 4 keywords from the most recent sentence added in the last of the queue clearly underlined. 
Users can also view previous context keywords using the previous/next arrow buttons.
Users select one or multiple context keyword(s) relevant to the content they want to record as a note (\eg``\textit{city}'' in Figure \ref{fig:flowchart} (b)), and the selected keywords are showcased on the right-hand side of the AR display. 
The selected keywords serve as quick notes, allowing users to express their thoughts and serve as reminders in a short period without recording complete sentences~\cite{chen2021exploring}.
Notably, only context keywords are displayed on the top section of the AR headset when users touch the button of the ring if no keyword is selected to prevent the AR contents from interfering with real-world activities.

\subsubsection{Customized Keyword Selection}

With the selected context keywords, the system can achieve within-context notes, as in~\cite{kalnikaite2012markup, haliburton2023walking}.
However, users often have their own note-taking habits and commonly-used words or phrases.
Therefore, users can incorporate customized keywords on the left-hand side of the AR display, which are pre-defined by the users in advance, appearing when a keyword is selected. 
The default customized keywords are WH words, including what, why and how, and a question mark ``?''. 
Users can select customized keywords to indicate whether a note is a question, and, if so, which type of WH or yes/no question it is (\eg``\textit{What}'' in \figname~\ref{fig:flowchart} (c)). 

\subsubsection{Derivative Keyword Selection}

Customized keyword selection involving users' habits and intentions might generate notes slightly beyond the context.
However, to indeed accomplish beyond-context notes, keywords other than context keywords are required.
In our brains, we usually think of something or some ideas when we are inspired by the things we see or hear, and we combine those things with our background knowledge, experiences and even personality to generate the thoughts.
Therefore, when taking a beyond-context note and cannot find proper context keywords, users can opt to double-click on the most relevant context keyword (\eg``\textit{rallies}'' in \figname~\ref{fig:flowchart} (d)) to generate derivative keywords via the LLM keyword derivation.
The original keyword and 4 derivative keywords, a total of 5 items, are presented in a circular layout at the center of the AR display. 
Users can select derivative keywords that best match their intentions (\eg``\textit{sign}'' in \figname~\ref{fig:flowchart} (e)), view more derivative keywords via the previous/next arrow buttons, or double-click on a derivative keyword close to their intentions to further obtain more relevant derivative keywords generated by the LLM.
Although the concepts of personal experiences and habits are used in customized keywords, these were not incorporated into derivative keywords. 
However, since users can further generate more relevant derivative keywords based on selected derivative keywords, even across multiple iterations if desired, this iterative process allows the newly generated derivative keywords to progressively narrow down and align more closely with users’ intentions.
Derivative keyword selection could take more time but obtain more precise notes matching users' intentions for beyond-context notes.
Therefore, it is a trade-off between time and precision for users.

\subsubsection{Candidate Sentence Selection}

The selected keywords could be from context, customized, and/or derivative keywords.
Once a keyword is selected, the system automatically composes 3 candidate sentences via the LLM sentence organization using the selected keyword(s) as input, positioned on the bottom section of the AR display. 
Whenever a new keyword is selected, new candidate sentences using all selected keywords are automatically generated (\figname~\ref{fig:flowchart} (b) (c) (e)).
Users can then select a candidate sentence that best matches their objectives as a note (\eg ``\textit{What signs were displayed...}'' in \figname~\ref{fig:flowchart} (f) (g)). 
However, if they cannot find any proper candidate sentence, they can double-click on any candidate sentence or selected keyword to skip the candidate sentence selection and directly record all selected keyword(s) (\eg``\textit{What, sign, city}'' in~\figname~\ref{fig:flowchart} (h) (upper)).
Furthermore, when they prefer not to spend excessive time and attention on note-taking and need to take a note hastily without reading candidate sentences or even waiting for sentence generation, they have the option to select only context (or also customized) keywords and record these as a \textit{quick note} through a double-click (\eg``\textit{What, city}'' in~\figname~\ref{fig:flowchart} (h) (lower) from only steps (a) (b) (c)).

\subsubsection{Note Refinement}

After taking notes, users can review their notes by clicking the ``Notes'' button, always on the display, as illustrated in~\figname~\ref{fig:refinement} (left).
When further selecting a note, they can see the transcripts of the selected keywords of the note, as in~\figname~\ref{fig:refinement} (right).
The note and transcripts as well as the keywords and candidate sentences are shown to remind users of the context at the time.
The history information also allows users to refine the note if they are not satisfied with it.
Previous/next arrow buttons are also for candidate sentences in refinement, so users can spend more time reading and finding the proper candidate sentence.
Note refinement is performed in a short time window that users might not be interested in the current topics of the activities, or after speeches before the Q\&A sessions, so users can swiftly organize or even perfect their notes.
Note refinement is actually not mainly for real-time note-taking, but such a swift note organization is performed right after speech-based activities and bridges the real-time note-taking and the complete note organization afterward.
Therefore, we still add this feature to make our system complete.

\begin{figure}
    \centering
    \includegraphics[width=1\linewidth]{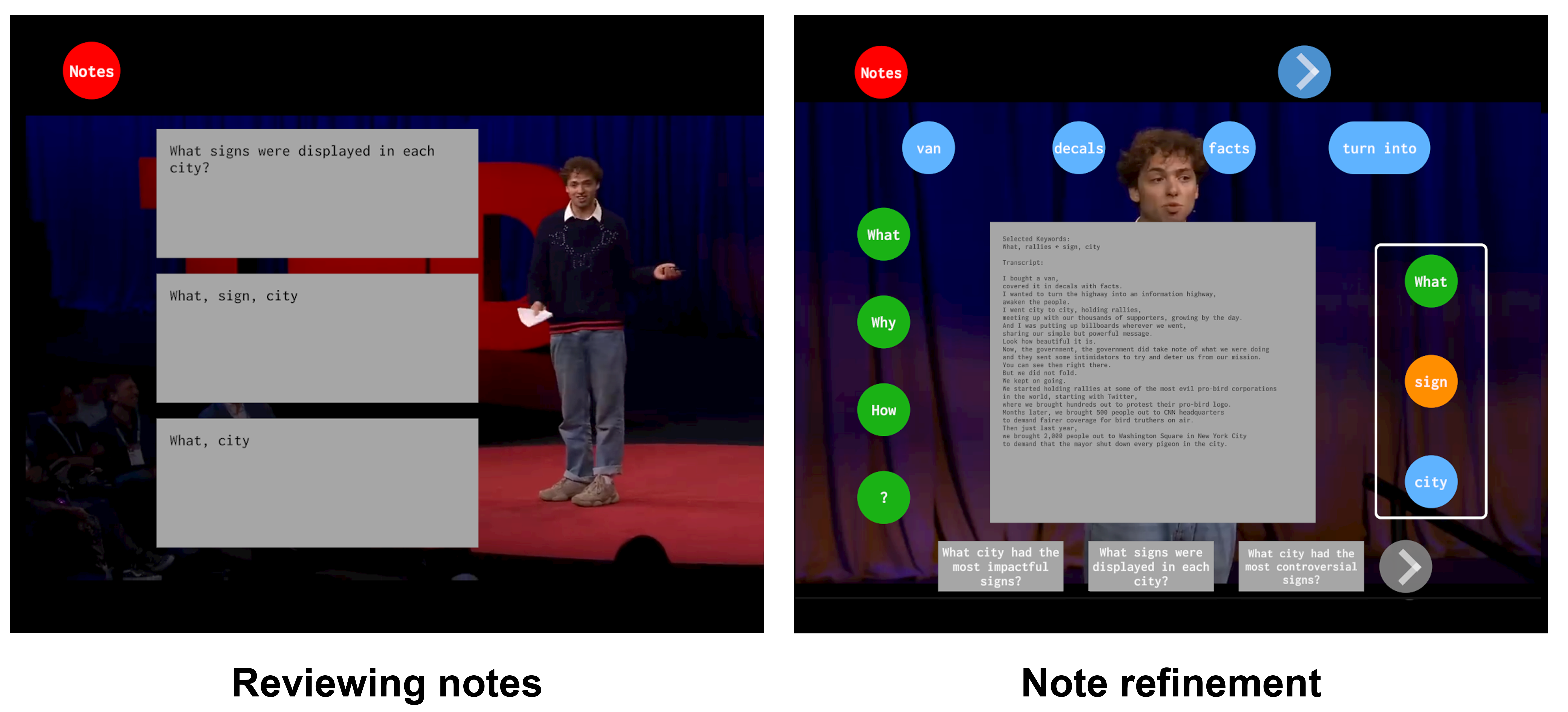}
    \caption{Reviewing notes by pressing the red ``Notes'' button and further reviewing transcripts and/or refining the notes.}
    \Description{(Left) Users could review the notes by pressing the red ''Notes'' button, always on the display. (Right) When further selecting a note, they can see the transcripts of the selected keywords of the note. The note and transcripts as well as the keywords and candidate sentences are shown to remind users of the context at the time.}
    \label{fig:refinement}
\end{figure}

In general, users primarily select context keywords and may also select a customized keyword(s), and then select a candidate sentence as a note or directly record selected keywords as a note.
Furthermore, they can select only context (or also customized) keywords to record these as a quick note in just two or three selection steps, which means that a \textit{quick note} is defined as one recorded only from (a) (b) or (a) (b) (c) steps in~\figname~\ref{fig:flowchart} in GazeNoter.
Performing a few keyword and sentence selection steps to take notes accomplishes the real-time system (DC1) and distraction and cognitive load (DC3) design considerations.
Using context keyword selection, customized keyword selection, and candidate sentence selection, the system matches users' intentions for within-context notes.
Advanced derivative selection is used only for beyond-context notes, and by incorporating this feature, the system fulfills the matching users' intentions design consideration (DC2).
Although taking notes while staying engaged in the ongoing activity could more or less cause distraction, GazeNoter allows users to freely employ their strategies to take notes based on the density of information in the speech-based activity. 
Instead of complete sentences, quick notes with only context (or also customized) keywords could be recorded in high-density information parts of the speech-based activity to avoid missing content.
Furthermore, two types of historical information are preserved, allowing users to review them in less stressed parts. 
First, all previous context keywords can be viewed using the previous/next
arrow buttons. 
Second, in note refinement, all selected keywords and their transcripts are also recorded. 
Therefore, users can add notes even if they have missed some content.
Notably, since notes automatically generated by current LLMs could be comprehensive without any miss and could be integrated in our system in the future, our goal is to precisely generate notes match users’ intentions for both within-context and beyond-context notes.

\subsection{AR Gaze Selection}

We leverage an AR headset as a medium to implement the proposed user-in-the-loop note-taking system due to its input and output manners for displaying and adjusting the LLM-generated content.
For output, AR headsets are always worn on the head and the displayed content is usually transparent on their see-through displays.
Therefore, users can still see through the AR content to notice real-world activities.
Moreover, by placing the AR content or items around the center or on the edges of the display and leaving the center blank, such a layout allows users to see the real world when the AR content is shown, as in~\cite{cai2023paraglassmenu, janaka2022paracentral}, which is adopted in our AR layout design (\figname~\ref{fig:flowchart} (right)).
In addition, users can easily turn the display on and off to reduce the interference from the displayed AR content.
Unlike looking down at smartphones, displays on AR headsets do not require users to change their head or even eye direction to look at AR content, which reduces the time and steps of switching between the displayed content and ongoing activities and is a subtle manner.
These benefits of AR headsets achieve swift, subtle, and low-distraction output for users in the real-time system (DC1), distraction and cognitive load (DC3), and subtlety (DC4) design considerations.

\begin{figure}
\begin{center} 
\includegraphics[width=0.75\linewidth]{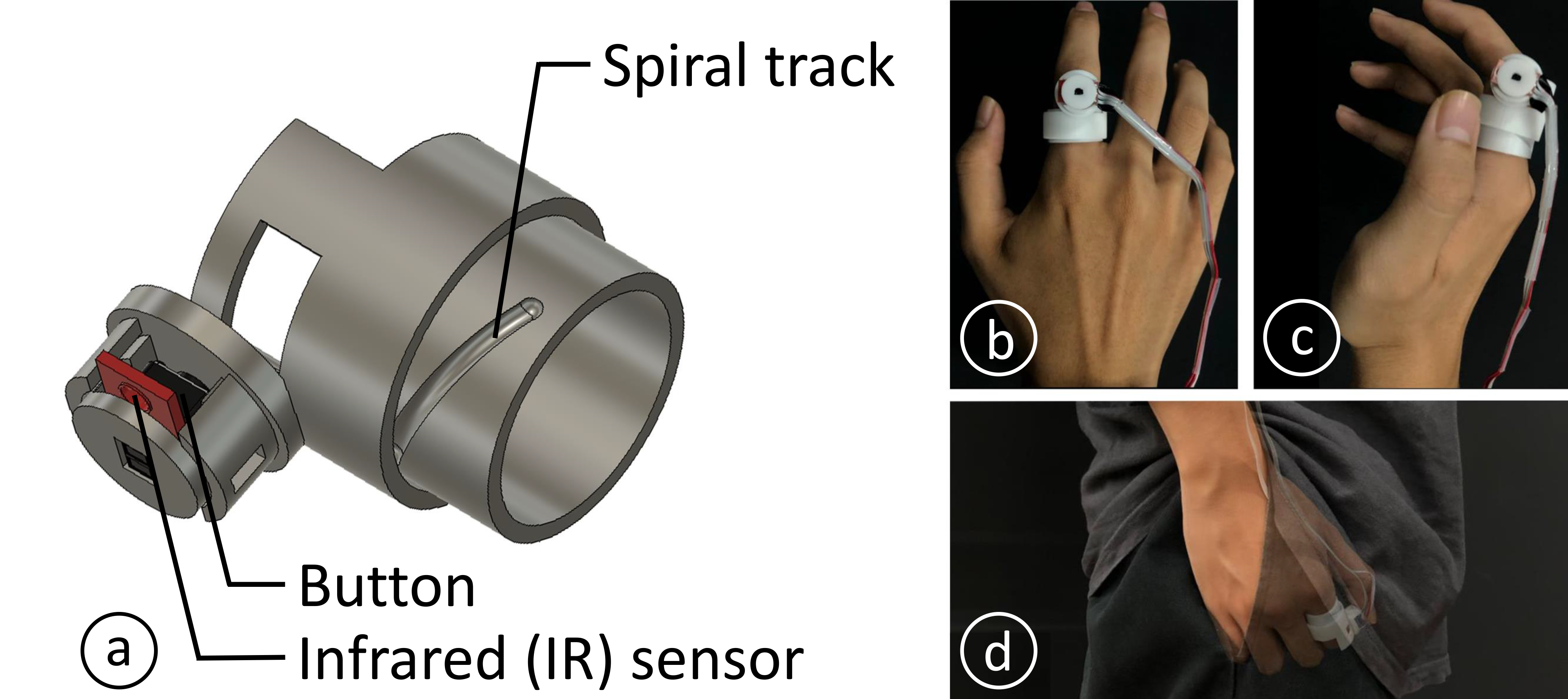}
\caption{(a) The hardware structure of the ring. (b) The button can be withdrawn on the back of the finger, preventing interference with users. (c) The button is extended for input. (d) The ring can be used subtly, such as in a pocket.}
\Description{(a) The hardware structure of the ring, consisting of a button, an infrared (IR)
sensor on top of the button and spiral tracks, worn on the index finger for confirmation. (b) The button can be withdrawn on the back of the finger, preventing interference with users. (c) The button is extended through the spiral track for input. (d) The ring can be used subtly, such as in a pocket.}
\label{fig:ring}
\vspace*{-6pt}
\end{center}
\end{figure}

For input, compared with gesture, controller and voice input, gaze selection is a more swift and subtle input approach, and the eye-tracking system is commonly built in off-the-shelf headsets.
Although several gaze confirmation methods are proposed, \eg dwelling and gaze gestures, the combination of gaze selection and gesture or ring confirmation has become popular and practical on AR headsets, such as Microsoft Hololens 2, Meta Quest Pro, Meta Orion and Apple Vision Pro. 
Therefore, we designed a ring, consisting of a button, an infrared (IR) sensor and spiral tracks, worn on the index finger for confirmation, as shown in~\figname~\ref{fig:ring}.
Users press the button with their thumb to confirm the gaze-selected item.
Users can even perform the confirmation using the ring when putting their hand in a pocket during discussions or conversations, accomplishing better subtlety and mobility.
This cannot be achieved by gesture input and confirmation due to the occlusion problem in gesture tracking.
The ring can be substituted by any advanced confirmation approaches in the future, such as picoRing~\cite{takahashi2024picoring}.
Besides, since 2D input on a small ring could suffer from a large control-display (CD) ratio issue, we chose the common combination of gaze selection and ring confirmation.
This combination fulfills not only swift (DC1) and subtle (DC4) but also low-distraction input (DC3) in the design considerations.

Besides gaze selection, rapidly turning the display on and off to achieve rapid switching between virtual and real worlds is another merit of AR headsets.
An IR sensor (QTR-1RC Reflectance Sensor) is equipped on the button of the ring, and a small case with a small hole is above the IR sensor.
When the thumb touches the button and covers the hole, the AR content is displayed, and the thumb can further press the button with the mechanical ``click'' feedback for confirmation.
Double-click is also used for the alternative confirmation, which means two consecutive clicks within 500ms. 
Once the thumb leaves the button, the display is turned off.
Therefore, The touch gesture via the IR sensor enables users to rapidly switch between AR content and real-world activities.
Notably, a solid background for the text of the AR content is used to prevent the font color from blending with the dynamically changing real-world colors.
Although a transparent background could minimize distraction, a solid background with the rapid switching design via the touch gesture reduces distraction and achieves a trade-off between distraction and the clarity and readability of the AR content.
Despite using a solid background, the AR layout leaves the center blank, as in ~\cite{cai2023paraglassmenu, janaka2022paracentral}.
Therefore, users can still see through the display.
With the rapid switching design on the ring, users can spend most of their time without AR content.
In addition, whenever the thumb touches the button, the AR content in the designed layout is anchored in the physical space relative to the headset direction at the initial moment of touching the ring.
Users can dynamically adjust the AR content position by re-touching the button when the real-world target is in the blank center of the AR layout.
As long as the thumb stays on the button, the AR layout remains fixed in the physical space.
Thus, users can select items close to the edges of the field of view (FoV) of the AR display easily by slightly moving the head. 
Notably, all AR items in the layout are within the AR display, so users can only move their eyes to achieve AR gaze selection, and the head movement is only auxiliary. 

Besides, the button should be on the back of the finger and the ring should be worn on the index finger close to the palm or the metacarpophalangeal (MCP) joints as a conventional ring to prevent it from interfering with users.
During using the system, the button should be moved to a comfortable location on the index finger to touch and press.
Therefore, the spiral tracks on the ring, as screw threads, are used to extend and rotate the button of $90^\circ$ to the position approximately at the proximal interphalangeal (PIP) joint, between the second and third finger segments, on the side of the index finger, which is in the comfort zone for touching with the thumb, based on~\cite{tsai2016thumbring}.
Users can move the button back after using the system, like wearing a conventional ring.
Such a withdrawable design is essential in AR and cross reality (XR) device design, as shown in~\cite{teng2021touch, tsai2022fingerx}.

Although gaze selection with ring confirmation has been explored, the two innovative features of the ring well-suited to our system include: (1) rapid switching between AR content and real-world activities via an IR sensor in a single touch-and-press step to reduce distraction and mitigate the blocking issue, and (2) AR content position adjustment achieved by touching the ring for mobile, real-time conditions. 
The proposed AR gaze selection with the ring achieves swift and subtle gaze selection, rapid switching across reality, AR content position adjustment, and conventional ring form factor.
While the performance of the input method was not evaluated since it is not the primary contribution of this paper, we still verified and discussed the merits and usability of the designs in both static and mobile note-taking scenarios in the following studies.
The button and IR sensor are controlled by a NodeMCU-32S microcontroller.
The weight of the 3D-printed ring and the sensors is 4.9g.
The lightweight ring and the AR headset, which could be envisioned to be as small as normal glasses in the future, in GazeNoter accomplish the mobility design consideration (DC5).

\subsection{Note-Taking System Implementation}

The details of system implementation and the three capabilities of the LLM for context keyword extraction, derivative keyword derivation, and candidate sentence organization are described in the following.
OpenAI (GPT-4) API is used to implement the LLM.
The prompts are listed in the appendix.
The flowchart of the system is illustrated in~\figname~\ref{fig:flowchart}.

\subsubsection{Context Keyword Extraction}
In the beginning, the audio speech is received by the headset and is then converted into text using the pre-trained model, Whisper, provided by OpenAI (size of the model: base, English-only model). 
Every 4 seconds we record a voice segment as the input to the Whisper model.
If there is a pause of over 1 second in records, it is treated as the start of a new sentence in the transcription.
The output transcript is used as input for the LLM to generate context keywords.
We observed that a sentence or part of a long sentence with a comma is usually said in 4 seconds, so this can generate context keywords of the sentence in real time.
The average delay for context keyword generation is about 4.29 seconds after users hear the sentence. 
In addition to the current sentence, previous sentences in the transcripts are also used as input for the LLM.
These provide context for the LLM to generate more precise context keywords of the sentence.
The LLM was prompted to extract at most 4 keywords from each sentence.
This number range was chosen based on a pilot study to avoid redundancy and ensure that essential keywords are included.
We originally designed a 2 $\times$ 5 layout to display 10 context keywords, enabling users to view context keywords from the two most recent simultaneously.
However, considering the guideline that the item size for gaze input should exceed $2^{\circ}$ in~\cite{kim2022lattice} and the precision of the eye-tracking system in off-the-shelves AR headsets, a 1 $\times$ 4 layout was finally adopted.
This design ensures that the item size is much larger than $2^{\circ}$ and the context keywords are placed sparser.

\subsubsection{Derivative Keyword Derivation} 

For beyond-context notes, users select the context keyword most relevant to their intentions and use it as input for the LLM to generate 4 derivative keywords, which are relevant to and derived from the context keyword.
This is like using the LLM as a brain to derive or associate relevant keywords from/with the context keyword for users.
4 derivative keywords consist of 2 types, including 2 derivative keywords \textit{contextually associated with the original keyword}, which are relevant to both the context and the original context keyword, and 2 derivative keywords \textit{exclusively related to the original keyword}, which are not relevant to the context but only relevant to the original context keyword.
The prompts for both types instruct the LLM to generate related words while prohibiting those overlapping with the displayed context keywords or having the same lemma.
This maximizes the number of new keywords present to the users.
The prompt for derivative keywords contextually associated with the original keyword includes the preceding 15 sentences as the context for the LLM.
We observed that 15 sentences can provide adequate context without including older topics unrelated to the current context based on a pilot study.
The prompt rule ensures that derivative keywords are related to the original keyword but remain distinct by avoiding overlap with current and past keywords and using different lemmas.
When users intend to further obtain more proper derivative keywords by double-clicking on a derivative keyword close to their intentions, the derivative keyword is used as input for LLM to generate more derivative keywords.
That original derivative keyword and the other 4 derivative keywords exclusively related to that original derivative keyword are shown in the circular layout.
We initially aimed to allocate 6 or even 9 derivative keywords, including the original keyword, in a circular layout, but 5 items in a circular layout enable easier and more accurate derivative keyword selection.
The average delay of derivative keyword derivation is about 1.41 seconds after double-clicking a keyword.

\subsubsection{Candidate Sentence Organization} 

Once a keyword is selected, the system automatically uses all selected keywords as input of the LLM to generate 3 candidate sentences. 
We prompt the LLM to ensure that each sentence either contains all selected keywords or shares the same lemma of words with the selected keywords.
Furthermore, the context of the previous transcripts is included.
We also limit each candidate sentence to 10 words for ease of reading.
Furthermore, if any WH words or a question mark from customized keywords is selected, the candidate sentences must be in the form of questions.
The average delay of candidate sentence generation is about 2.89 seconds.
The layout of 3 candidate sentences on the bottom section of AR display is based on the design in~\cite{choi2022kuiper}, aiming to make it more comfortable for users to read the sentences, which require longer reading time.

\section{User Study 1: Formal Speech}

To understand how users leverage the GazeNoter system, verify its usability, and compare it with other note-taking methods in a static speech scenario, we conducted this user study.
Notably, instead of assessing user comprehension or memory retention in note-taking, our primary goal was to evaluate the practical utility and gather subjective feedback on the note-taking experience with GazeNoter.
Therefore, we followed the established methodologies for evaluating novel interaction techniques and systems, especially for LLM interactions ~\cite{jiang2023graphologue, suh2023sensecape} and note-taking ~\cite{Chandrasegaran2019talktraces, fang2021notecostruct, haliburton2023walking}, in this user study.

\subsection{Participants and Apparatus}

An Oculus Quest Pro headset with a built-in eye-tracking system was used. 
5 ring sizes, ranging from 18mm to 22mm in diameter, were prepared.
The ring for gaze confirmation was worn on the index finger.
A smartphone, iPhone 8 plus (5.5-inch display), was used for comparison. 
12 participants (5 female) aged 19-27 (mean: 24.16) were recruited. 
They were compensated with 10 USD for their time.

\begin{figure}
\begin{center} 
\includegraphics[width=0.9\linewidth]{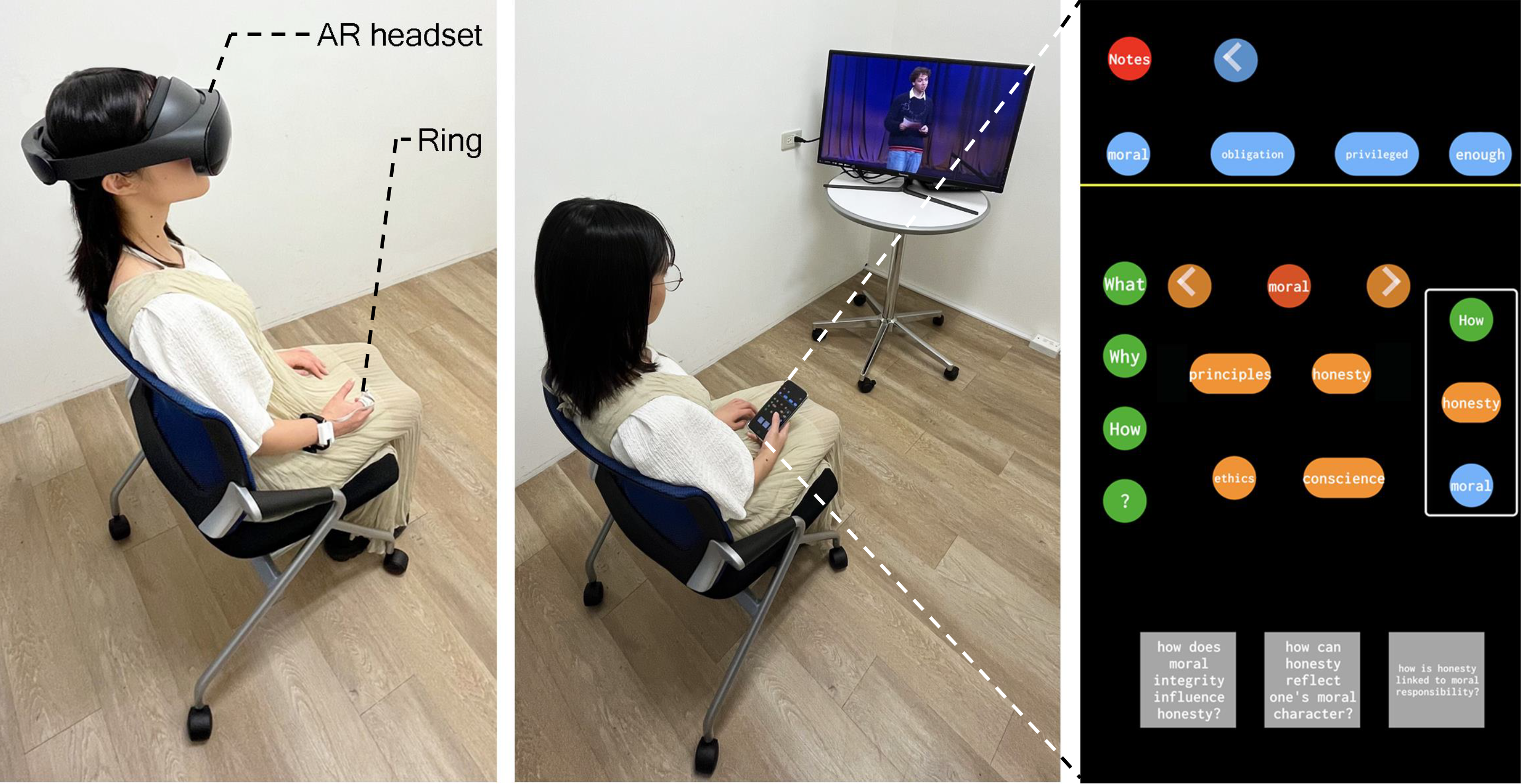}
\caption{Study 1 setup of GazeNoter on the AR headset (G) (left) and the smartphone (S) (middle) with its layout (right).}
\Description{Study 1 setup of GazeNoter on the AR headset (G) (left) and on the smartphone (S) (middle) with its layout on the smartphone (right).}
\label{fig:setup}
\vspace*{-6pt}
\end{center}
\end{figure}

\subsection{Task and Procedure} \label{Task and Procedure}

\subsubsection{Task}
To simulate attending speeches, 6 TED Talks videos with similar difficulty and length, about 10 minutes, were used.
The TED Talks are similar to lecture scenarios that were extensively studied in prior research~\cite{mosleh2016challenges, piolat2005cognitive} but under more time constraints for note-taking.
Three note-taking methods were compared, including smartphone text entry (T), smartphone version GazeNoter (S), and the proposed AR GazeNoter (G).
An iPhone built-in note-taking app, Notes, was used in the smartphone text entry (T), which was used as a baseline. 
For the smartphone version GazeNoter (S), we implemented the proposed system on the smartphone.
All contents were shown on the smartphone screen, and the layout was similar to that on the AR display but was slightly adjusted to fit in the commonly-used portrait mode of the smartphone (\figname~\ref{fig:setup}).
The font size was the same as in the built-in note-taking app in (T).
Since the blank center was not needed on the smartphone, the circular layout for derivative keywords was not used.
Touch input was used to select items on the smartphone via tap and double-tap gestures.
By comparing (T) and (S) on the smartphone, we could understand whether our users-in-the-loop LLM system outperformed the current manual text entry note-taking methods.
By comparing (S) and (G), we could observe the distraction, cognitive load subtlety, and social acceptance of GazeNoter on the smartphone and AR headset. 
(S) and (G) are both methods proposed in this paper but on different media.
Notably, we also intended to observe whether the notes from the proposed user-in-the-loop LLM system outperformed automatically generated notes by the LLM.
Therefore, notes were automatically generated by using the whole speech transcript for the LLM after the speech in (G).
This was an additional automatic note-taking method (A) as another baseline although no user input was in this method.
The generated note number in (A) was the same as in (G) by each participant.
Thus, this additional comparison was only between (A) and (G).

\subsubsection{Procedure}
Participants sat on a chair and listened to speeches.
After the experiment and GazeNoter were introduced, an eye-tracking calibration on the AR headset was performed, and a proper ring was worn. 
The speech videos were played on the AR display in a fixed space in the real world in (G) to simulate real speeches and on a monitor (32-inch display) on the desk in front of them in (T) and (S).
To prevent the noise in the environment from interfering with the performance, the audio source of the videos was directly received from the computer for speech-to-text conversion.
One video was used in a training session for all participants. 
We encouraged them to explore every feature and asked them to think about their quick note-taking strategies for GazeNoter based on their experiences in the training session.
During the experiment, they watched a TED Talks video in each method and performed note-taking for questions in the Q\&A sessions or important points and ideas they wanted to record.
After the speeches, they could review the notes, see the transcripts of the notes, or even perform refinement in a 30-second time window, as a small time window between a speech and the Q\&A session.
Certainly, these could be performed during the speeches.

\subsubsection{Measures}
The order of the methods was counterbalanced.
Besides one video for the training, one of the other five videos was randomly played in a method.
The time and steps of each note and each feature of GazeNoter, and the number of notes, keywords and sentences users selected were recorded.
Furthermore, participants needed to fill out a questionnaire with a 7-point Likert scale.
11 questions were for each method.
4 of them, including intention, reminding, quality and inspiration, were for notes from each method, and 4 were based on NASA-TLX, including cognitive load, frustration, physical effort and usability, as shown in~\figname~\ref{fig:likert}.
The other 3, including distraction, subtlety and social acceptance, were specifically added to understand participants' experiences when using these three methods to perform note-taking.
In quality, it only represents the quality of the notes but was irrelevant to whether the notes matched users' intentions, which was rated in intention.
In reminding, it means whether the notes from a method could remind users of their thoughts or opinions.
The 4 questions for notes were used to evaluate not only the overall ratings but also each note for deeper investigation.
Notably, since no user input was in (A), and the additional comparison was only between (A) and (G), participants only scored the 4 questions for the notes in (A).
They were interviewed for some feedback.
The study lasted approximately two hours, including the introduction, training sessions, breaks between methods and interview. 

\subsection{Result and Discussion} 

The results are shown in~\figname~\ref{fig:result} to~\ref{fig:likert}.
Repeated measures ANOVA and Bonferroni correction for post-hoc pairwise tests were used to analyze the objective, quantitative data, including time and numbers of notes and keywords.
A Friedman test and Wilcoxon signed-rank tests with Bonferroni correction for post-hoc pairwise tests were utilized to analyze the results of the subjective questionnaire. 

\subsubsection{Quantitative Results}
Objective and quantitative results showed significant effects in the number of notes ($F_{2, 22} = 14.08, p < 0.001$), time per note ($F_{2, 22} = 5.41, p = 0.01$), and keywords per note ($F_{2, 22} = 29.55, p < 0.001$). 
Post-hoc tests indicated significant differences in note quantities between (T, S) and (T, G), in note time for (T, G), and in keywords across all pairs. 
Despite (T)'s shorter note times, (S) and (G) had higher note counts and (G) had more keywords per note, suggesting a preference for more detailed notes despite the additional time required.

The time of a note was measured from the first selection of a keyword to the last selection of recording the note in (S) and (G), and from typing the first letter to pressing the last ``return'' button to complete the note in (T).
The number of keywords per note was calculated by including notes with only keywords and excluding notes with complete sentences. 
Furthermore, the percentage of notes with complete sentences is 28\% in (T), 73\% in (S), and 78\% in (G), and the others are notes with only keywords.
The percentage of quick notes is 21\% in (S) and 18\% in (G).
Quick notes were the notes recorded with only context (or also customized) keywords and completed with less than 2s in the final recording step via double-click, which was less than the average candidate sentence generation time of 2.89s. 
This condition indicates that participants intended to take notes quickly by selecting keywords without considering reading candidate sentences, which means only (a) (b) or (a) (b) (c) steps in~\figname~\ref{fig:flowchart}.
The average time of a quick note in (S) is 5.22s for context keywords only and 6.59s for customized keywords included.
The average time of a quick note in (G) is 5.05s for context keywords only and 5.85s for customized keywords included.
These demonstrate how fast the proposed system could be if quick notes are required.
Furthermore, by comparing steps and time of (G) in \figname~\ref{fig:S1G2Step} and~\ref{fig:S1G2 time}, we obtain that 4 seconds could allow the selection of about 2 to 4 context keywords in a note.
The percentage of beyond-context notes, meaning notes with derivative keywords selected, is 22\% in (S) and 28\% in (G) (\figname~\ref{fig:Appendix-S1G1}).
This shows the importance of beyond-context notes in note-taking.
Furthermore, since derivative keywords in (G) could cause more occlusion on the AR headset, it is noteworthy that derivative keywords are displayed for only 27\% of the total note-taking time. The percentage during the entire speech duration is even lower, which indicates that distraction is not severe.

\begin{figure}
\begin{center} 
\includegraphics[width=1\linewidth]{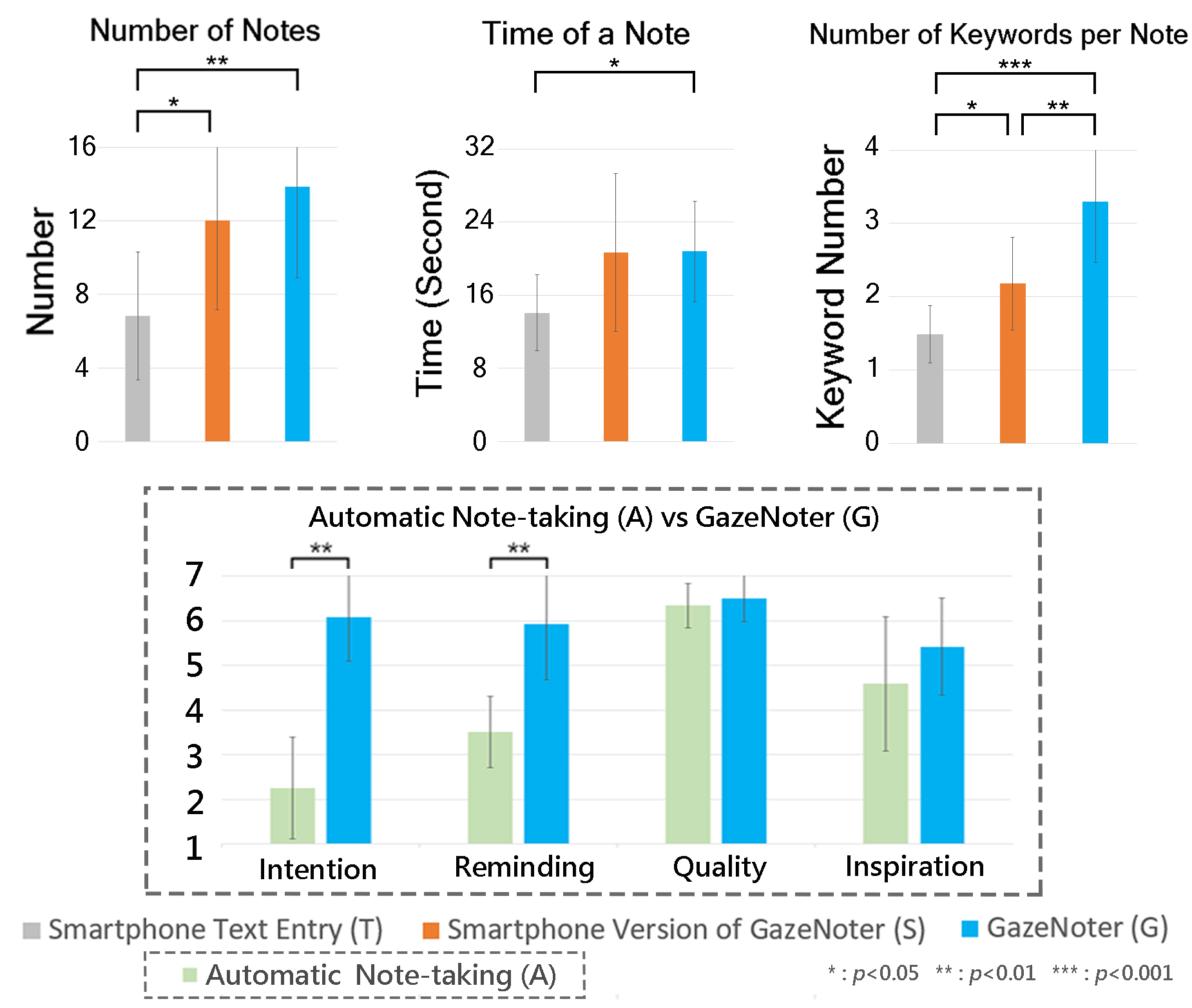}
\caption{Study 1 quantitative results of the three methods, (T), (S) and (G). 
The dashed line represents the qualitative results of the additional comparison between (A) and (G).
}
\Description{Study 1 quantitative results of the three methods, (T), (S) and (G). Showing (G) outperforms (S) and (T) in note quantitative and quality. The dashed line represents the qualitative results of the additional comparison between (A) and (G), illustrating that note outcome of (G) excels in terms of matching user intention and serves as a more effective reminder.}
\label{fig:result}
\vspace*{-6pt}
\end{center}
\end{figure}

\subsubsection{Subjective Scale Results}
For the results of the subjective questionnaire,
significant main effects are revealed in intention ($\chi^2(2) = 11.52, p < 0.01$), reminding ($\chi^2(2) = 10.09, p < 0.01$), quality ($\chi^2(2) = 21.41, p < 0.001$), inspiration ($\chi^2(2) = 19.00, p < 0.001$), 
distraction ($\chi^2(2) = 16.33, p < 0.001$), cognitive load ($\chi^2(2) = 12.05, p < 0.01$), frustration ($\chi^2(2) = 13.18, p < 0.01$), physical effort ($\chi^2(2) = 13.47, p < 0.01$), subtlety ($\chi^2(2) = 12.00, p < 0.01$), and usability ($\chi^2(2) = 17.91, p < 0.001$).
However, no significant main effect is found in social acceptance ($\chi^2(2) = 1.23, p = 0.53$).
Post-hoc pairwise tests show significant differences among all pairs in intention, usability, distraction and cognitive load, between (T, S) and (T, G) in reminding, quality, inspiration frustration and physical effort, and between (T, G) and (S, G) in subtlety.
(G) significantly outperforms (T) in most factors except for social acceptance, and (S) significantly outperforms (T) in most factors except for social acceptance and subtlety.
This indicates that the proposed user-in-the-loop AI system in (S) and (G) significantly improves note-taking in the baseline (T).
Furthermore, (G) significantly outperforms (S) in intention, subtlety, usability, distraction, and cognitive load indicating that leveraging an AR headset as a medium for the proposed user-in-the-loop AI system significantly enhances the note-taking experience.
Regarding the additional comparison between (A) and (G), significant main effects are revealed in intention ($\chi^2(1) = 12.00, p < 0.01$) and reminding ($\chi^2(1) = 12.00, p < 0.01$).
However, no significant main effects are found in quality ($\chi^2(1) = 2.00, p = 0.15$) and inspiration ($\chi^2(1) = 3.60, p = 0.05$).
This verifies the necessity of integrating user input in the user-in-the-loop AI system to match users' intentions and effectively remind them.
This verifies the necessity of integrating user input in the user-in-the-loop AI system to match users' intentions and effectively remind them.

\begin{figure}
\begin{center} 
\includegraphics[width=1\linewidth]{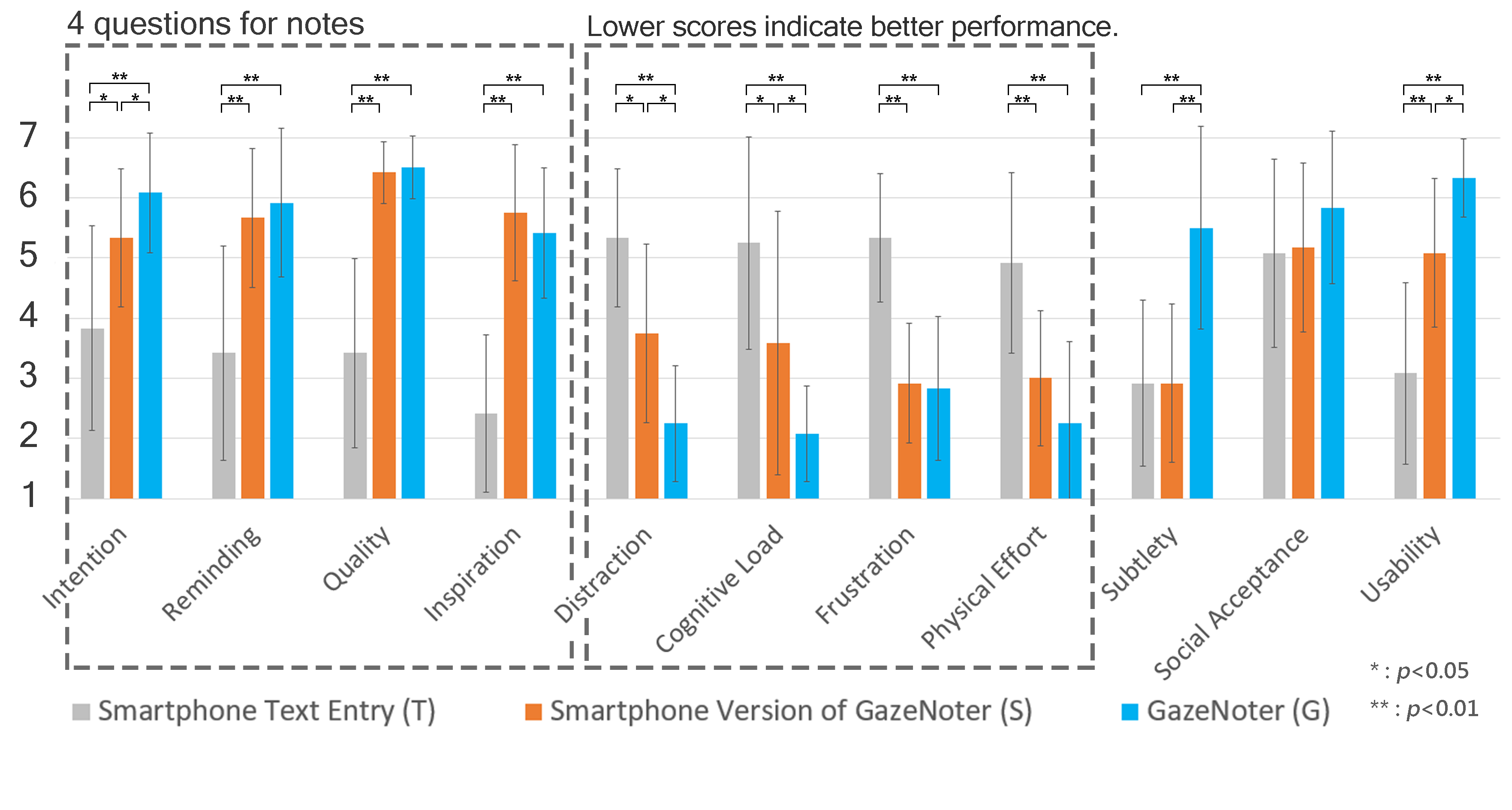}
\caption{Study 1 qualitative and subjective scale of the user study on a 7-point Likert scale. 
The left part highlighted by a dashed line represents the 4 questions for notes. 
The middle part highlighted by a dashed line represents the scales where lower scores indicate better performance.}
\Description{Study 1 qualitative and subjective scale of the user study on a 7-point Likert scale. The dashed line on the left highlights the 4 questions for notes from each method. The dashed line on the right highlights the scales that lower scores indicate better performance. Results show that (G) significantly outperforms (S) and (T) in various aspects, including intention, distraction, cognitive load, and usability.}
\label{fig:likert}
\vspace*{-6pt}
\end{center}
\end{figure}

\subsubsection{Additional Discussions}
\paragraph{AR Display}
In the objective and quantitative results, \textit{P1}, \textit{P4}, \textit{P7}, \textit{P8}, and \textit{P10} commented that context keywords were easy to access and organized, which encouraged them to take more notes.
\textit{P1}, \textit{P7} and \textit{P9} appreciated the LLM-generated sentence feature and were more willing to take notes due to no need to manually type.
For the time of a note and number of keywords per note, all participants reported that the AR content did not disrupt their viewing of the speech, except when momentarily using the derivative keyword feature, which required minimal time. 
7 participants (\textit{P3}, \textit{P5}, \textit{P7}, \textit{P8}, \textit{P9}, \textit{P10}, \textit{P11}) mentioned that they were accustomed to keeping the AR display on throughout the entire speech to show context keywords, which facilitated quicker and easier keyword selection.
\textit{P7} elaborated that \textit{``Always displaying the AR content allowed me to select the context keywords and sentences as soon as they were extracted and generated, respectively. 
However, achieving this on the smartphone (S) required constantly checking the screen, which was very distracting and hindered me from performing more keyword and sentence selection.''}
This is consistent with the findings in the note composition distribution in \figname~\ref{fig:Appendix-S1G1} of the appendix, where half of the notes in (S) consist solely of context keywords, requiring minimal time and steps.

Furthermore, \textit{P3}, \textit{P7}, \textit{P10} and \textit{P11} highlighted the seamless transition between engaging with the speech and note-taking on the AR headset in (G). 
\textit{P10} mentioned that \textit{``I could take my time to take notes, switch my focus from the AR content to the speech for a while to stay engaged, and then switch back for note-taking.''}
The seamless transition in (G) enhanced the flexibility of note-taking and seemed to increase the time of a note due to the inclusion of time spent engaging with the speech.
For (T), \textit{P3}, \textit{P11} and \textit{P12} mentioned recording minimal keywords due to the slow typing speed, resulting in a shorter note-taking time.
These could contribute that the time in (S) and (G) are similar but the number of keywords in (G) is significantly higher than the others.
This outcome is also supported by the data presented in \figname~\ref{fig:S1G2Step} and~\ref{fig:S1G2 time} in the appendix, where the time distribution for (S) and (G) is similar but more steps are in (G), indicating more efficient note-taking.

\paragraph{4 Questions for Notes}
All participants commented that upon selecting desired keywords, at least one of the generated candidate sentences would match their intentions.
This observation could be supported by the results that the significantly higher number of keywords per note in (G) leads to its intention score also being significantly higher than the others.
For within-context notes, most participants except for \textit{P6} and \textit{P8} could select desired context and customized keywords. 
The average intention score from each within-context note, as rated by participants, is 6.5. 
However, \textit{P6} and \textit{P8} encountered a situation where the desired keywords were heard but not extracted as context keywords, diminishing the relevance between the notes and their intentions. 
For beyond-context notes, half participants (\textit{P1}, \textit{P3}, \textit{P4}, \textit{P5}, \textit{P7}, \textit{P9}) successfully selected desired keywords, subsequently recording an adequate candidate sentence as a note.
Nevertheless, the others struggled to always select desired derivative keywords, resulting in less relevant notes. 
As a result, although the average score of each note on intention for beyond-context notes remains high at around 6 points, it is still slightly lower than that for within-context notes (\figname~\ref{fig:S1EachNote}).
For the additional subjective scales, the 4 questions for each note in~\figname~\ref{fig:S1EachNote}, statistical analysis was not performed since the number of within-context and beyond-context notes are different.
The scores of beyond-context notes are similar to or slightly lower than those of within-context notes. 
This indicates that participants were still generally satisfied with the beyond-context notes.

In terms of reminding and quality, all participants believed that sentences could more effectively remind them than keywords, so they would prefer to record sentences as notes in (S) and (G) if sentences aligned with their intention. 
Furthermore, they all agreed that the LLM-generated sentences were easy to understand and well-structured.
On the contrary, 7 participants (\textit{P1}, \textit{P2}, \textit{P3}, \textit{P4}, \textit{P5}, \textit{P9}, \textit{P12}) mentioned that they tended to type only fragmented sentences or a few keywords in (T) due to the slow typing process diverting their attention. 
These fragmented notes did not effectively remind them and had poor grammar. 
Regarding inspiration, \textit{P1}, \textit{P3}, \textit{P7}, and \textit{P10} emphasized that sentences often acted as a source of inspiration in (S) and (G). 
When reviewing the notes, the notes not only reminded them but also sparked new ideas or innovative perspectives. 
\textit{P3} further detailed that \textit{``The LLM-generated sentences were sometimes rephrased or restructured to elaborate the concept mentioned by the speaker, thereby fostering new insights.''}
These result in both (S) and (G) significantly outperform (T) in reminding, quality, and inspiration. 
For the additional comparison between (A) and (G), all participants agreed that the automatically LLM-generated notes misaligned with their intentions, and lacked an effective reminding feature, primarily due to their unfamiliarity with these notes.
This underlined the necessity of the user-in-the-loop AI system to match users' intentions.
The abovementioned merits of LLM-generated content were in both (A) and (G).
\textit{P3} and \textit{P8} commented that automatically LLM-generated notes helped them to discover new insights they had missed, which was inspiring.

\paragraph{Subjective Task Loads}
Regarding distraction and cognitive load, all participants agreed that maintaining visual attention on the ongoing speech while taking notes was a clear advantage of using the AR headset in (G). 
This reduced distraction and cognitive load caused by missing out on the speech content.
\textit{P7} and \textit{P10} stated that they usually read all candidate sentences in (G) to create more complete and longer notes, which could even inspire new ideas.
However, \textit{P5} and \textit{P8} mentioned that the learning curve for gaze selection could initially increase their cognitive load, but it would become highly efficient and convenient once they were familiar with it.
\textit{P5} and \textit{P9} further specified that the gaze selection in (G) prevented them from selecting one keyword while looking at another, which is an inherent limitation of gaze selection and required time to adapt.
For (T), all participants reported that the slow typing speed and typos led to high cognitive load and even frustration.
\textit{P1}, \textit{P6} and \textit{P10} further commented that the need to structure the notes, even with minimal keywords, greatly increased their cognitive load.
These contribute to significantly higher distraction, cognitive load, and frustration in (T).

In terms of subtlety, 6 participants (\textit{P2}, \textit{P4}, \textit{P9}, \textit{P10}, \textit{P11}, \textit{P12}) agreed that (G) only requiring minimal eye and finger movement made it imperceptible.
On the contrary, 7 participants (\textit{P1}, \textit{P2}, \textit{P4}, \textit{P6}, \textit{P10}, \textit{P11}, \textit{P12}) pointed out that looking down at the smartphone was quite noticeable and tiring in both (T) and (S), especially in (T) due to the longer duration for manual typing.
These result in (G) being the most subtle method with the least physical effort.
For social acceptance, most participants commented that taking notes during a speech was common and would not be considered impolite, even if constantly looking at the smartphone.
This causes no statistical significance in social acceptance.
For usability, all participants appreciated the context keyword feature in (G) as automatic highlights of the speeches, which was not achieved in (S) due to the need to look down the smartphone.
Furthermore, \textit{P4}, \textit{P8}, \textit{P9} and \textit{P10} found the selection feature in (S) and (G) instead of manual typing to be very useful.
\textit{P10} elaborated that \textit{``This combination of AR and AI overcomes all the shortcomings of real-time note-taking.''}
These result in significantly better usability in (G).
For note refinement, only \textit{P2} and \textit{P10} selected two notes recorded with only keywords to review the transcripts but did not modify or refine the notes, respectively.
Besides, extractive note-taking, which typically involves directly copying, highlighting or quoting the original text, and abstractive note-taking, which generally involves rephrasing or summarizing based on the original context, are two critical concepts in note-taking.
Since sentences of notes are generated by the LLM based on the selected keywords, the GazeNoter system focus on achieving abstractive note-taking for both within-context and beyond-context notes. 
However, since transcripts are also recorded in note refinement, extractive note-taking is inherently accomplished as well. 
Based on the results of 4 questions for notes, and with only \textit{P2} and \textit{P10} reviewing the extractive notes in note refinement, users expressed greater satisfaction with the abstractive notes from GazeNoter compared to the extractive notes.

In general, the proposed method, GazeNoter (G), significantly outperforms the others in various accepts.
The LLM-generated notes from (S), (A) and (G) have better quality and inspiration than the manual typing notes from (T).
The user-in-the-loop LLM system in (S) and (G) makes the notes better match users' intentions and remind them than the automatic LLM-generated notes from (A) and even the manual typing notes from (T).
The proposed keyword and sentence selection for note-taking in (S) and (G) outperforms manual typing in (T), and using the AR headset as a medium for the selection with minimal eye and finger movement in (G) further results in less distraction and cognitive load, as well as better subtlety and usability compared to using the smartphone in (S).
While the three methods have similar social acceptance, this is caused by the speech scenario.
Therefore, we evaluated the face-to-face discussion scenario in the following study.

\section{User Study 2: Walking Meeting}

In the previous study, the performance of GazeNoter in the speech scenario was evaluated in a static condition with less interference.
We further intended to evaluate its performance in a mobile condition, so attending walking meetings, an emerging use case and research area~\cite{damen2020hub,damen2020understanding, haliburton2023walking}, was the scenario examined in this study.
Moreover, walking meetings involve interactive communication and occasional face-to-face discussion, resulting in frequent eye contact and less time for note-taking.
Hence, the mental pressure in this scenario could be higher compared to simply listening to speeches in the previous study.

\begin{figure}
\begin{center} 
\includegraphics[width=0.75\linewidth]{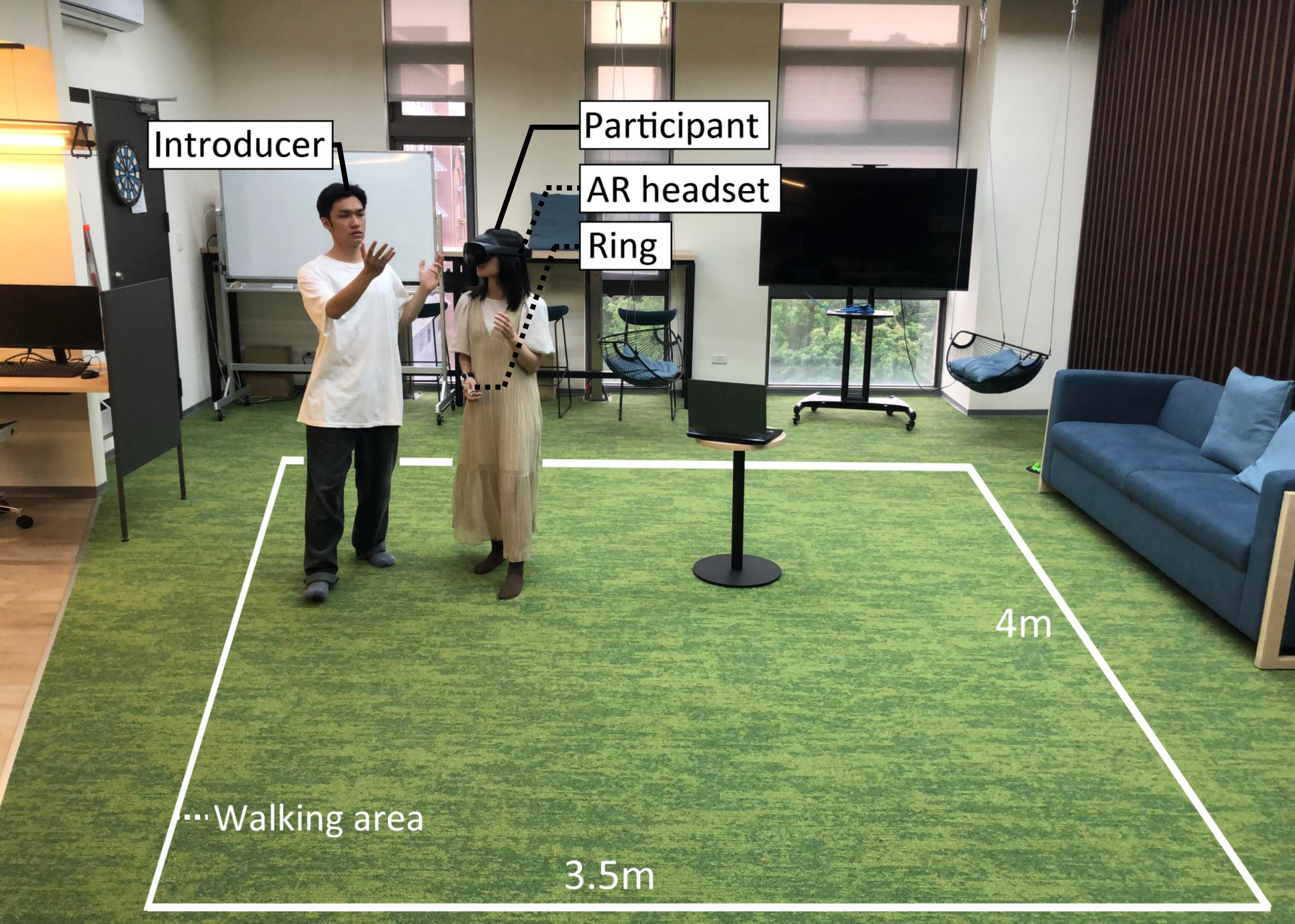}
\caption{Study 2 setup for walking meetings in an area measuring 3.5m $\times$ 4m with an introducer. 
}
\Description{Study 2 setup for walking meetings in an area measuring 3.5m $\times$ 4m with an introducer.}
\label{fig:casestudy}
\vspace*{-6pt}
\end{center}
\end{figure}

\subsection{Setup, Task and Procedure}

The setup was similar to the previous study.
12 participants (6 female) aged
22-27 (mean: 25.08), who had not participated in the prior study, were recruited for this study.
Participants were compensated with 10 USD for their time.
Instead of sitting on a chair, an introducer and a participant walked around a room measuring 3.5m $\times$ 4m for walking meetings.
The introducer initiated a topic and spent 3 to 5 minutes describing it, followed by a discussion with the participant during walking meetings. 
Each walking meeting was about 10 minutes.
The participant could freely take notes throughout the description and discussion.
However, the participant might need to review the notes for discussion with the introducer.
This increased the mental pressure and required more urgent real-time note-taking.
As in the previous study, three methods (T), (S) and (G) were examined in each walking meeting, and an additional comparison between (A) and (G) was performed.
Similarly, the order of the methods was counterbalanced, and participants filled out a questionnaire and were interviewed after the experiment.
The study took about two hours. 

\subsection{Results and Discussion} 

\begin{figure}
\begin{center} 
\includegraphics[width=1\linewidth]{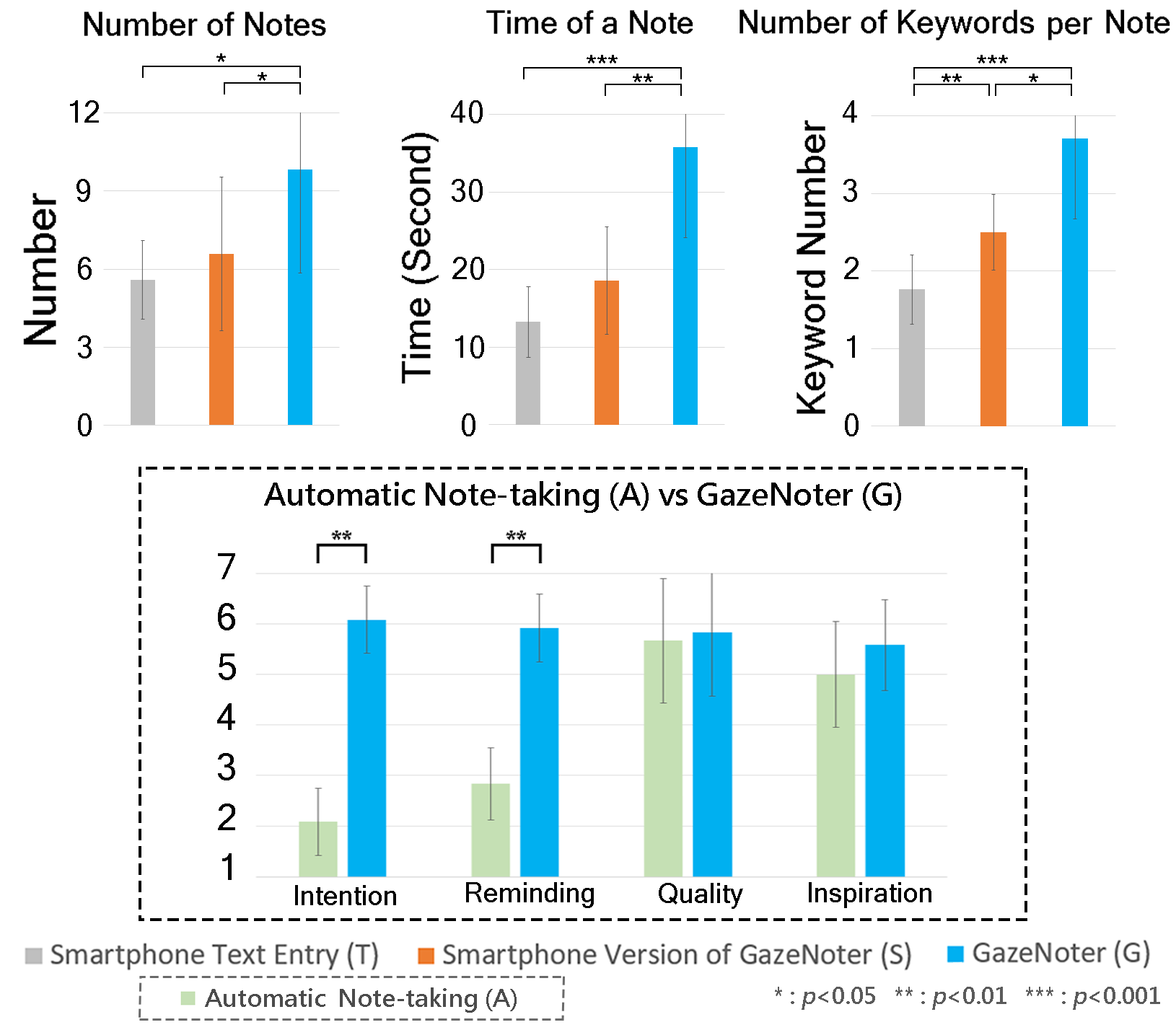}
\caption{Study 2 quantitative results of the three methods, (T), (S) and (G). The dashed line represents the qualitative results of the additional comparison between (A) and (G).}
\Description{Study 2 quantitative results of the three methods, (T), (S) and (G). Showing (G) outperforms (S) and (T) in note quantitative and quality. Furthermore, participants were willing to spend more time taking notes in dynamic conditions when using (G). The dashed line represents the qualitative results of the additional comparison between (A) and (G), illustrating that note outcome of (G) excels in terms of matching user intention and serves as a more effective reminder.}
\label{fig:result2}
\vspace*{-6pt}
\end{center}
\end{figure}

The results are shown in~\figname~\ref{fig:result2} to~\ref{fig:likert2}.
Repeated measures ANOVA and Bonferroni correction for post-hoc pairwise tests were used to analyze the objective, quantitative data.
A Friedman test and Wilcoxon signed-rank tests with Bonferroni correction for post-hoc pairwise tests were utilized to analyze the subjective questionnaire. 

\subsubsection{Quantitative Results}

For the objective and quantitative results, significant main effects are revealed in number of notes ($F_{2, 22} = 9.81, p < 0.01$), time of a note ($F_{2, 22} = 26, p < 0.001$) and number of keywords per note ($F_{2, 22} = 16.97, p < 0.001$).
Post-hoc pairwise tests show significant differences between (T, G) and (S, G) in the number of notes and time of a note, and among all pairs in the number of keywords per note.
The percentage of notes with complete sentences is 23\% in (T), 37\% in (S), and 63\% in (G).
The quick note percentage is 42\% in (S) and 26\% in (G) as shown in the note type distribution of \figname~\ref{fig:Appendix-S2G1} in the appendix.
The average time of a quick note in (S) is 5.96s for context keywords only and 7.45s for customized keywords included.
The average time of a quick note in (G) is 4.76s for context keywords only and 6.26s for customized keywords included.
This scenario leads participants to spend less time on note-taking and prefer quick notes compared to the previous study.
For the results of the subjective questionnaire,
The beyond-context note percentage is 14\% in (S) and 18\% in (G) (\figname~\ref{fig:Appendix-S2G1}).
The derivative keywords are displayed 13\% of the total note-taking time in (G).

\subsubsection{Subjective Scale Results}
For the results of the subjective questionnaire, significant main effects are revealed in intention ($\chi^2(2) = 22.14, p < 0.001$), reminding ($\chi^2(2) = 15.95, p < 0.001$), quality ($\chi^2(2) = 18.57, p < 0.001$), inspiration ($\chi^2(2) = 13.35, p < 0.01$), 
distraction ($\chi^2(2) = 22.37, p < 0.001$), cognitive load ($\chi^2(2) = 14.63, p < 0.01$), frustration ($\chi^2(2) = 14.00, p < 0.01$), physical effort ($\chi^2(2) = 17.15, p < 0.001$), subtlety ($\chi^2(2) = 22.14, p < 0.001$), social acceptance ($\chi^2(2) = 17.66, p < 0.01$), and usability ($\chi^2(2) = 21.14, p < 0.001$).
Post-hoc pairwise tests show significant differences among all pairs in all factors, except between (S, T) in social acceptance.
The additional subjective scales of the 4 questions for each note are shown in~\figname~\ref{fig:S2EachNote}.
The differences among the methods are more apparent in this scenario.
For the additional comparison between (A) and (G), significant main effects are revealed in intention ($\chi^2(1) = 12.00, p < 0.01$) and reminding ($\chi^2(1) = 12.00, p < 0.01$), and no significant main effects are found in quality ($\chi^2(1) = 0.14, p = 0.70$) and inspiration ($\chi^2(1) = 2.67, p = 0.10$), consistent with the previous study.

\begin{figure}
\begin{center} 
\includegraphics[width=1\linewidth]{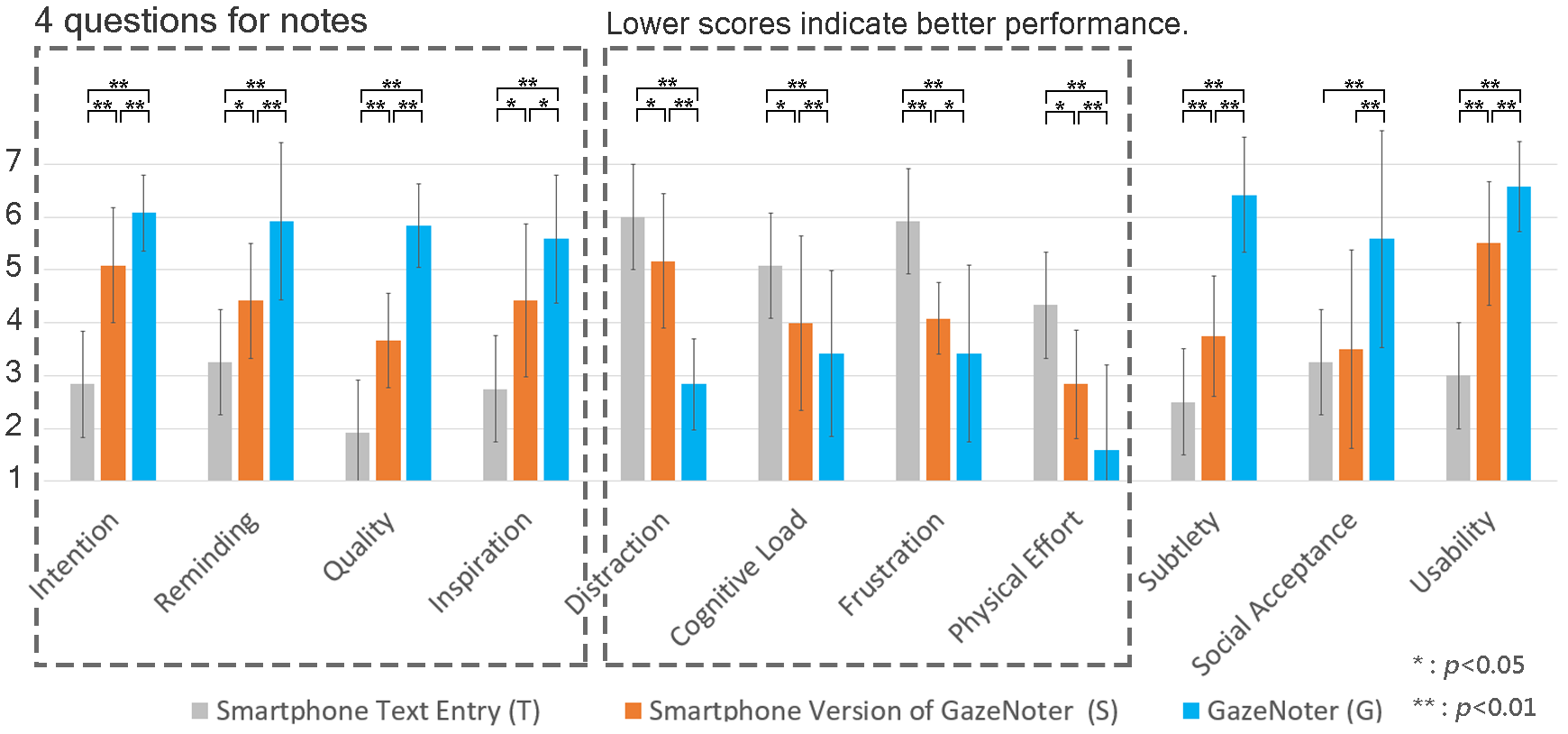}
\caption{Study 2 qualitative and subjective scale of the user study on a 7-point Likert scale. 
The left part highlighted by a dashed line represents the 4 questions for notes. 
The middle part highlighted by a dashed line represents the scales where lower scores indicate better performance.}
\Description{Study 2 qualitative and subjective scale of the user study on a 7-point Likert scale. The dashed line on the left highlights the 4 questions for notes from each method. The dashed line on the right highlights the scales that lower scores indicate better performance. Results show that (G) significantly outperforms (S) and (T) in various accepts, including intention, reminding, quality, inspiration, distraction, cognitive load, frustration, physical effort, subtlety, and usability.}
\label{fig:likert2}
\vspace*{-6pt}
\end{center}
\end{figure}

\subsubsection{Additional Discussions}
Similar feedback and comments were mentioned for several factors as in the previous.
We focus on discussing the differences caused by the mobile condition, interactive communication, and occasional face-to-face discussion in the walking meeting scenario.
(G) significantly outperforms both (T) and (S) in the number of notes and time of a note.
6 participants (\textit{P2}, \textit{P4}, \textit{P9}, \textit{P10}, \textit{P11}, \textit{P12}) specified that the diverted attention caused by looking down at the phone in (T) and (S) during discussions led them to take fewer notes and spend less time on each note.
This is also shown in \figname~\ref{fig:S2G2 step} and~\ref{fig:S2G2 time}, demonstrating that the time users spent on each note composition was consistently longer in (G) compared to (S).

For social acceptance, \textit{P10} mentioned the necessity of seeking permission to take notes in (S) and (T) to avoid being perceived as impolite.
\textit{P7} pointed out that they had to pause the discussion and sometimes even stop walking to take notes in (S) and (T), which hindered the ongoing conversation.
As for (G), two users (\textit{P2} and \textit{P4}) perceived the necessity to slightly move their eyes for interaction during face-to-face conversation diminishing social acceptance while others found it natural to glance away occasionally, enabling them to take notes.
These also result in significantly higher social acceptance in (G) compared to (T) and (S), and underscore the influence of social settings on note-taking during occasional face-to-face discussions in walking meetings.
Regarding subtlety and physical effort, \textit{P1}, \textit{P4}, \textit{P9}, and \textit{P10} specified that constantly switching attention among the walking path, the phone, and the introducer with frequent head movement in (T) and (S) was physically demanding and less subtle. 
Thus, (G) significantly outperforms (T) and (S) in physical effort and subtlety.
In addition, the longer duration for manual typing in (T) results in (T) being significantly less subtle than (S) in walking meetings.

For frustration, 6 participants (\textit{P1}, \textit{P3}, \textit{P4}, \textit{P7}, \textit{P10}, \textit{P11}) supposed that 
it was challenging to read text on the smartphone in (T) and (S) in the mobile condition of walking.
\textit{P4} further specified that \textit{``During walking, it was hard to maintain focus on the smartphone, read the texts and select keywords or sentences (S), not to mention organizing thoughts and then manually typing the notes (T).''  }
Therefore, (G) significantly outperforms both (T) and (S) in frustration.
\textit{P7}, \textit{P10}, and \textit{P11} mentioned that they gave up candidate sentence selection and directly recorded keywords as quick notes for less reading time. 
Consequently, the percentage of notes with complete sentences drops from 73\% to 37\% and the percentage of quick notes rises from 21\% to 42\% in (S). 
These disparities in (S) are much more pronounced compared to (T) and (G), as shown in the note type distribution of \figname~\ref{fig:Appendix-S2G1} in the appendix.
This also leads to significantly better performance in reminding, quality, and inspiration in (G) compared to (S).
For (G), although the percentage of notes with complete sentences also drops from 78\% to 63\% and the percentage of quick notes rises from 18\% to 26\% (\figname~\ref{fig:Appendix-S2G1}), this difference is not as pronounced as in (S).
\textit{P2}, \textit{P7} and \textit{P9} mentioned that they would opt to record all selected keywords when they lacked the time to read the candidate sentences, such as in high-density information part of the speech-based activity.
\textit{P1} and \textit{P5} also commented that they could obtain the desired sentences in notes when only selecting and recording keywords during such a situation.
This demonstrates that participants can dynamically adapt their note-taking strategies based on different situations, balancing the trade-off between creating more complete and detailed longer notes and producing more concise and quicker notes.
For note refinement, no participants used this feature due to the interactive communication and mobile condition in this scenario.

In general, GazeNoter (G) significantly outperforms the other methods in all aspects.
The requirement for occasional face-to-face discussions and the mobile condition in walking meetings in this study result in obvious performance differences between GazeNoter on the smartphone (S) and the AR headset (G).
Compared to smartphones, the absence of the need to look down, rapid switching across reality, and the large display are the primary benefits of AR headsets causing (G) to significantly outperform (S) in this study.

\section{Limitations and Future Work}

In general, GazeNoter was appreciated by participants in both studies.
However, there are still some limitations.
The precision of the built-in eye-tracking system in commercial AR headsets could occasionally disrupt the selection process, frustrating users. 
Using an advanced eye-tracking system in the future could improve not only frustration but also subtlety and social acceptance by reducing the time spent on diverted attention during selection. 
Another limitation is that real-world object or human tracking is not incorporated into the current system.
Therefore, the current AR layout is not dynamically and automatically placed around the speaker, as in ParaGlassMenu~\cite{cai2023paraglassmenu}.
This will be achieved in the future work.
Furthermore, blurring and increasing transparency of items not selected or gazed at could reduce interference from AR content. 
However, the dynamic change and delays in these adjustments could interfere with users due to the rapid movement of the gaze.

While our system applies personal experiences and habits to customized keywords, it does not extend this to derivative keywords. To generate beyond-context notes more tailored to individual users, incorporating personal knowledge, experiences, and personality when creating derivative keywords could be beneficial. 
However, while tailored generation offers advantages, balancing user-specific outputs with out-of-the-box inspiration will be essential.
The token limit of the LLM should be considered as well.
Certainly, typing input could achieve the most precise notes.
However, due to the real-time system design consideration (DC1), typing was not included in our design.
The results of the studies also show that GazeNoter on both the smartphone and AR headset outperforms smartphone text entry.
Nevertheless, using LLM to generate or auto-complete notes based on typed keywords is an advanced combination of the benefits of typing and our system, which generates notes (candidate sentences) based on selected keywords.
Typing could be more precise but may cause more distraction.
Therefore, comparing this and our system would be interesting in the future.

Although users can freely employ their note-taking strategies and two types of historical information are preserved, including previous context keywords in context keyword selection and transcripts in note refinement, to prevent missed content, we also envision that notes automatically generated by the LLM, as the baseline (A) in the studies, could be incorporated in the future.
Although these notes may not always match users’ intentions and are limited to within-context, they can provide comprehensive contextual coverage to compensate for potential missed content. 
Thus, automatic note-taking (A) is complementary to our GazeNoter system.

\section{Conclusion}

In this paper, we propose a real-time note-taking system, GazeNoter, by integrating a user-in-the-loop LLM system with gaze selection on an AR headset to generate notes that are both within-context and beyond-context, matching the users' intentions. 
We evaluated GazeNoter's effectiveness in two different scenarios: the static one where the user is stationary and listening to a speech, and the mobile one where the user is walking and participating in a meeting. 
Our findings indicate that GazeNoter, through simple keyword and sentence selection processes on either smartphones or the AR headset, significantly outperforms both manual note-taking and automatic notes generated by the LLM in many metrics. 
Furthermore, when using the AR headset, GazeNoter demands less eye, head and hand movement, leading to less distraction and cognitive strain. 
More broadly, our work contributes to the ongoing efforts to integrate foundation models into XR applications.

\begin{acks}
This research was supported in part by National Science and Technology Council of Taiwan (NSTC 111-2221-E-004 -008 -MY3, 113-2221-E-004 -007) and National Chengchi University.
\end{acks}

\bibliographystyle{ACM-Reference-Format}
\bibliography{sample-base}
\appendix
\section{Additional objective result}
\newpage

\begin{figure}[H]
    \centering
    \includegraphics[width=0.9\linewidth]{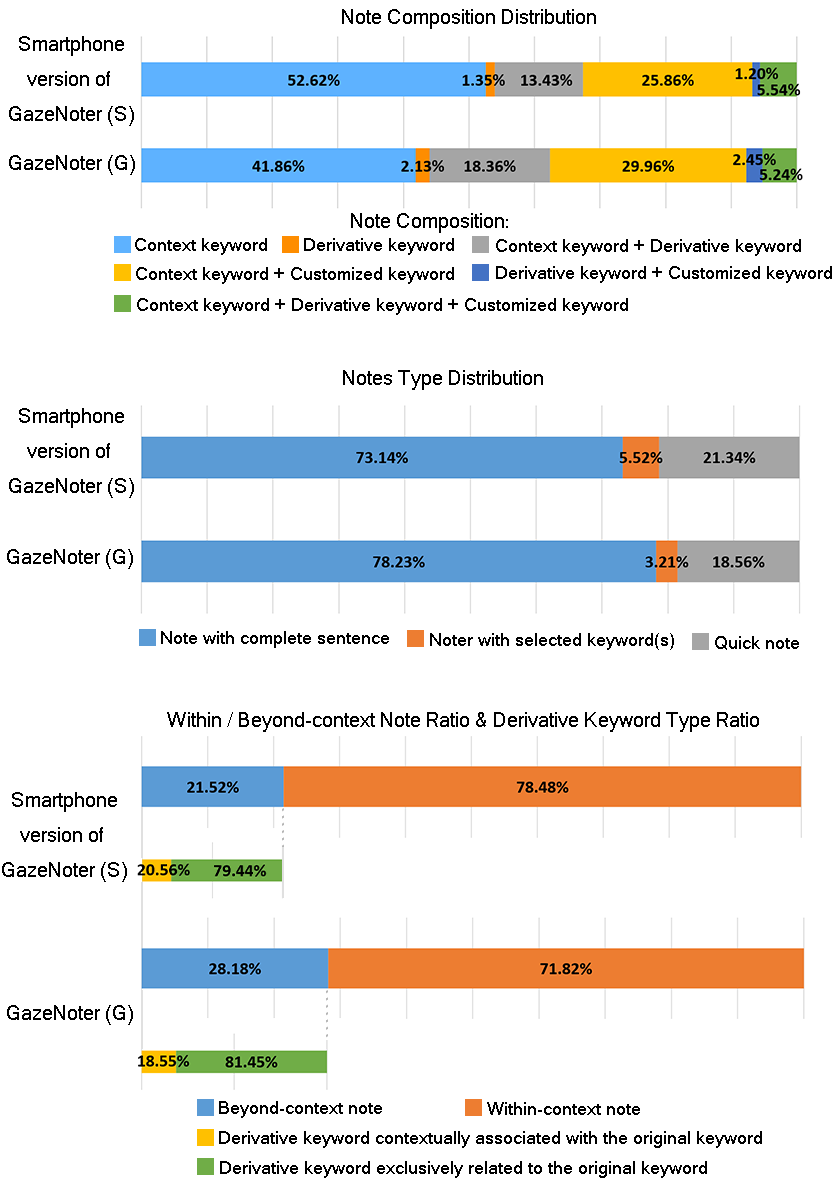}
    \caption{Objective results of the note structures comparison between (S) and (G) in user study 1.}
    \Description{Objective results of the note structures comparison between (S) and (G) in user study 1.}
    \label{fig:Appendix-S1G1}
\end{figure}

\begin{figure}[H]
    \centering
    \includegraphics[width=0.75\linewidth]{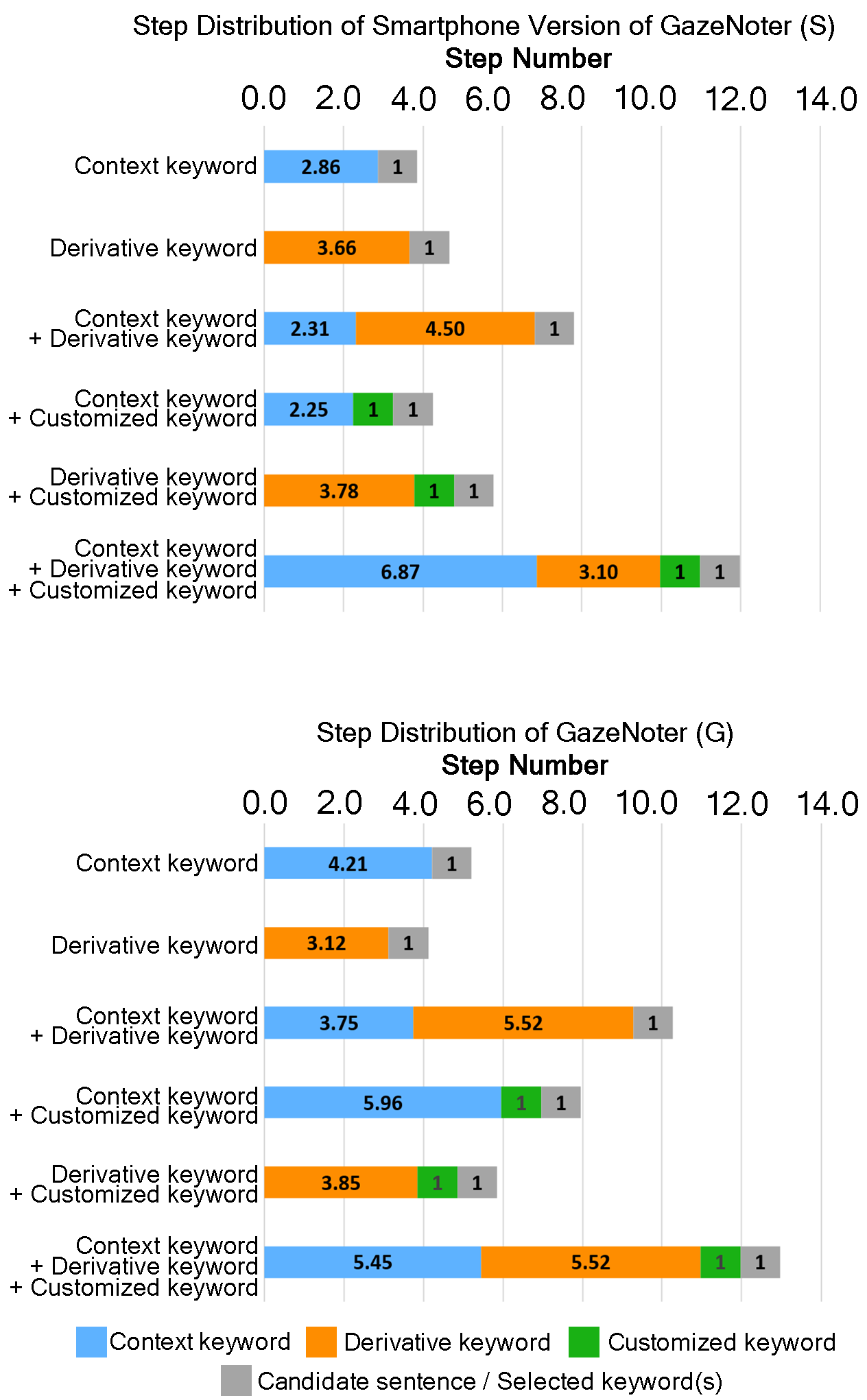}
    \caption{The step distribution of note compositions of (S) and (G) in user study 1.}
    \Description{The step distribution of note compositions of (S) and (G) in user study 1.}
    \label{fig:S1G2Step}
\end{figure}

\begin{figure}[H]
    \centering
    \includegraphics[width=0.75\linewidth]{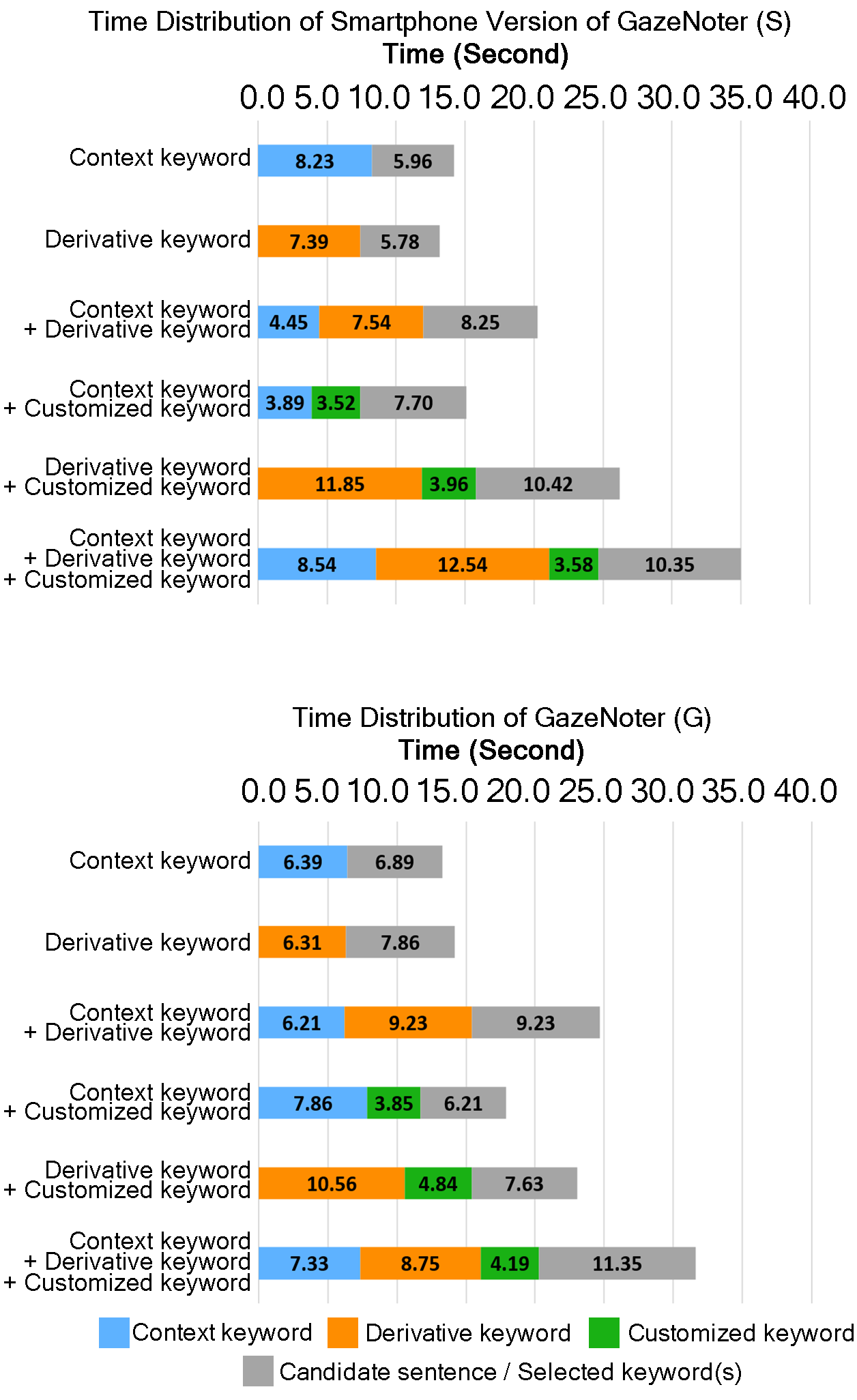}
    \caption{The time distribution of note compositions of (S) and (G) in user study 1.}
    \Description{The time distribution of note compositions of (S) and (G) in user study 1.}
    \label{fig:S1G2 time}
\end{figure}

\begin{figure}[H]
    \centering
    \includegraphics[width=0.75\linewidth]{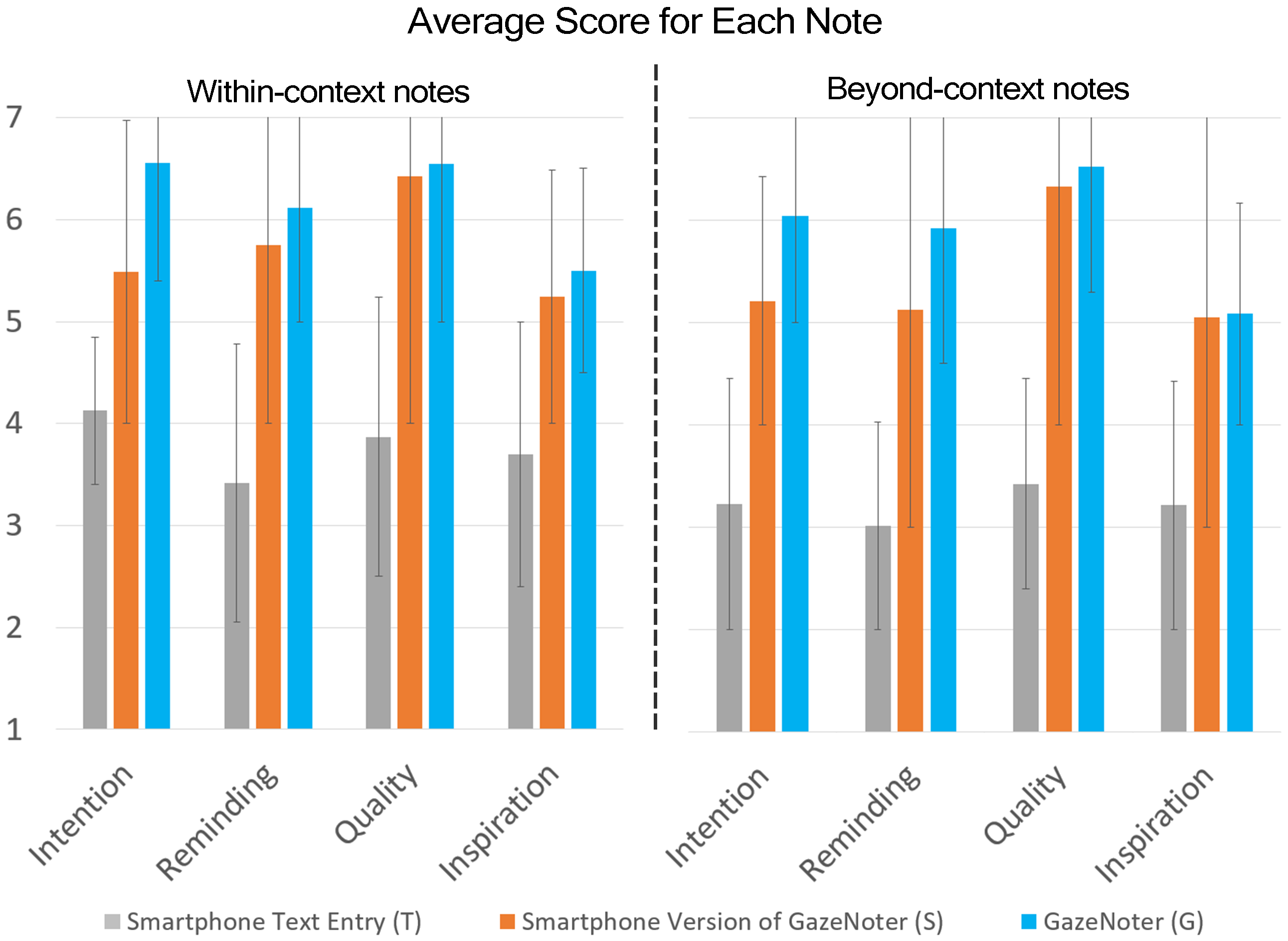}
    \caption{Score from each within-context and beyond-context note in user study 1.}
    \Description{Score from each within-context and beyond-context note in user study 1.}
    \label{fig:S1EachNote}
\end{figure}

\begin{figure}[H]
    \centering
    \includegraphics[width=0.9\linewidth]{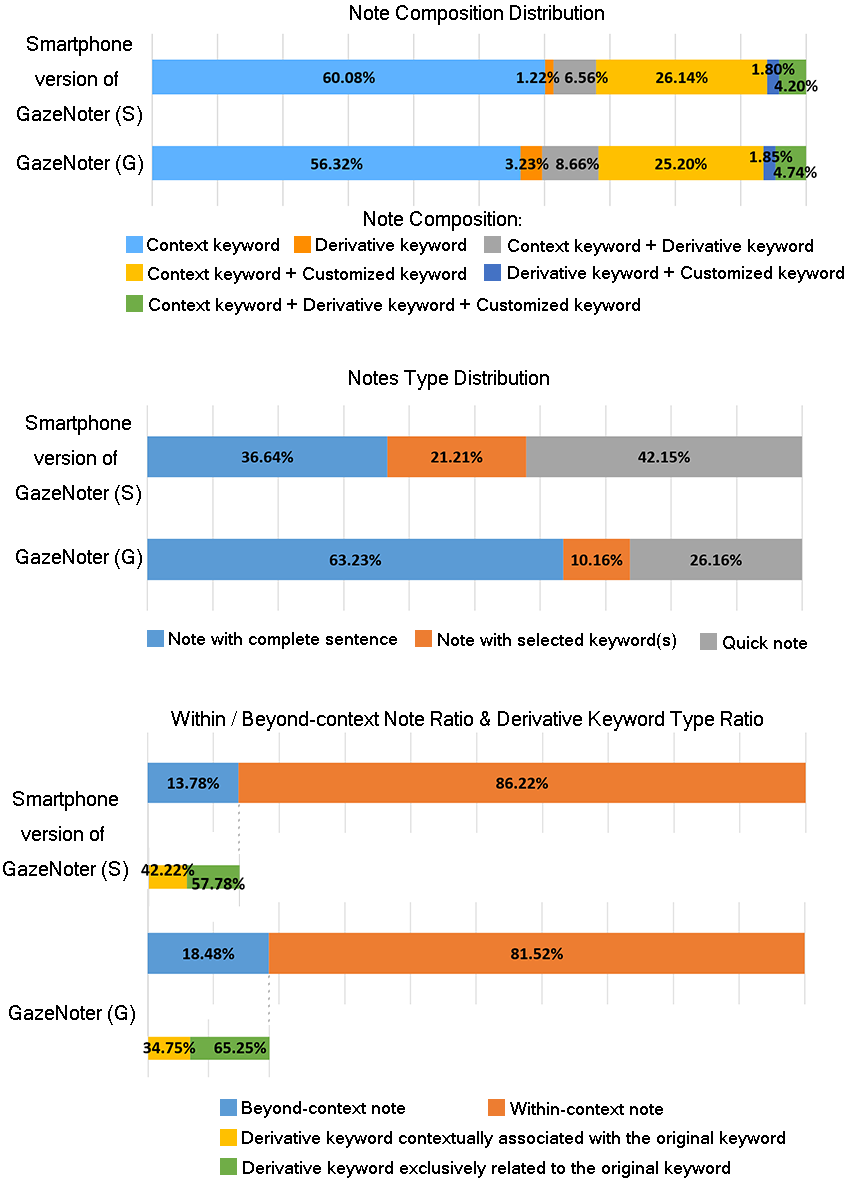}
    \caption{Objective results of the note structures comparison between (S) and (G) in user study 2.}
    \Description{Objective results of the note structures comparison between (S) and (G) in user study 2.}
    \label{fig:Appendix-S2G1}   
\end{figure}

\begin{figure}[H]
    \centering
    \includegraphics[width=0.8\linewidth]{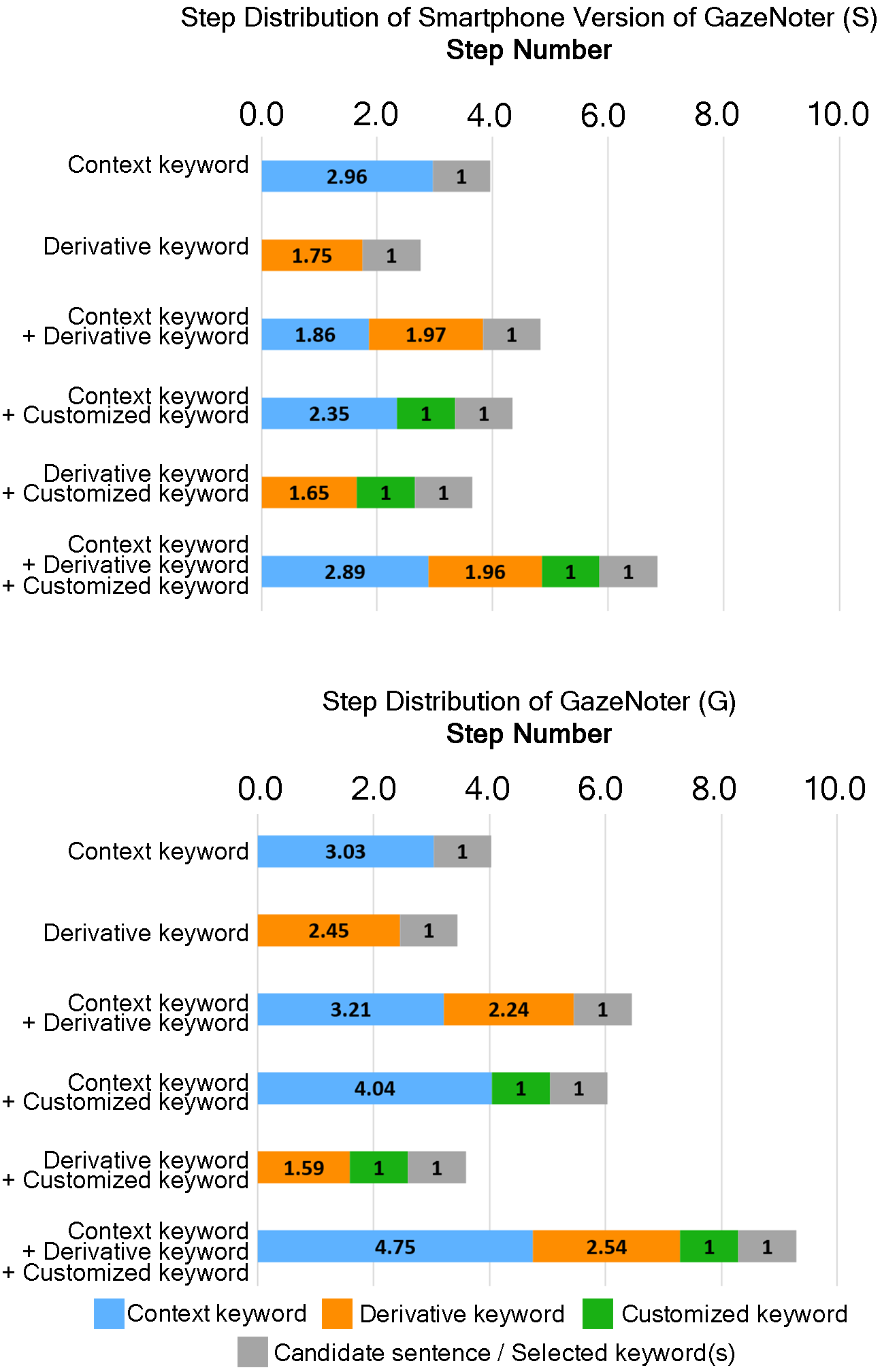}
    \caption{The time distribution of note compositions of (S) and (G) in user study 2.}
    \Description{The time distribution of note compositions of (S) and (G) in user study 2.}
    \label{fig:S2G2 step}
\end{figure}

\begin{figure}[H]
    \centering
    \includegraphics[width=0.8\linewidth]{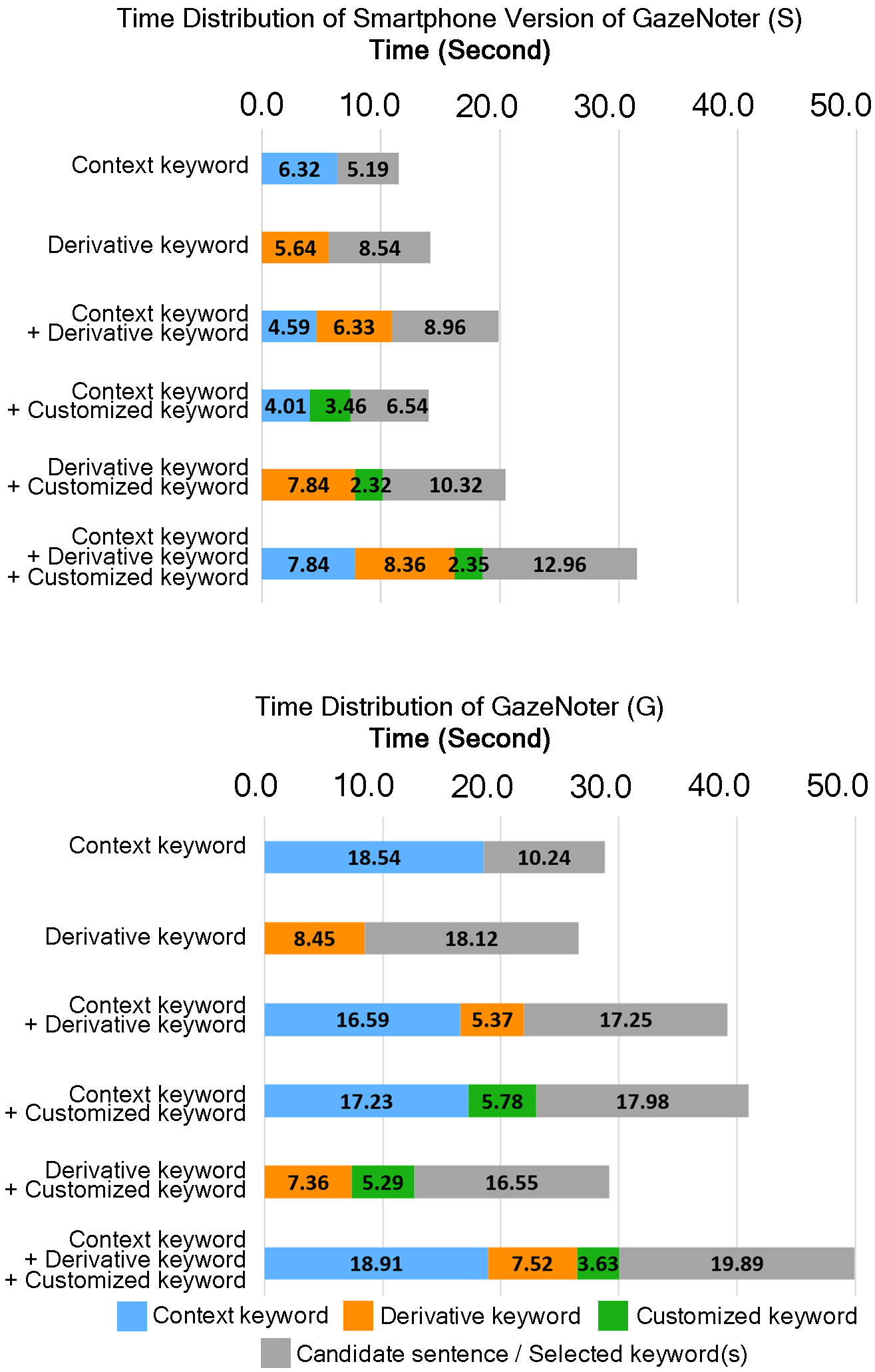}
    \caption{The time distribution of note compositions of (S) and (G) in user study 2.}
    \Description{The time distribution of note compositions of (S) and (G) in user study 2.}
    \label{fig:S2G2 time}
\end{figure}

\begin{figure}[H]
    \centering
    \includegraphics[width=0.75\linewidth]{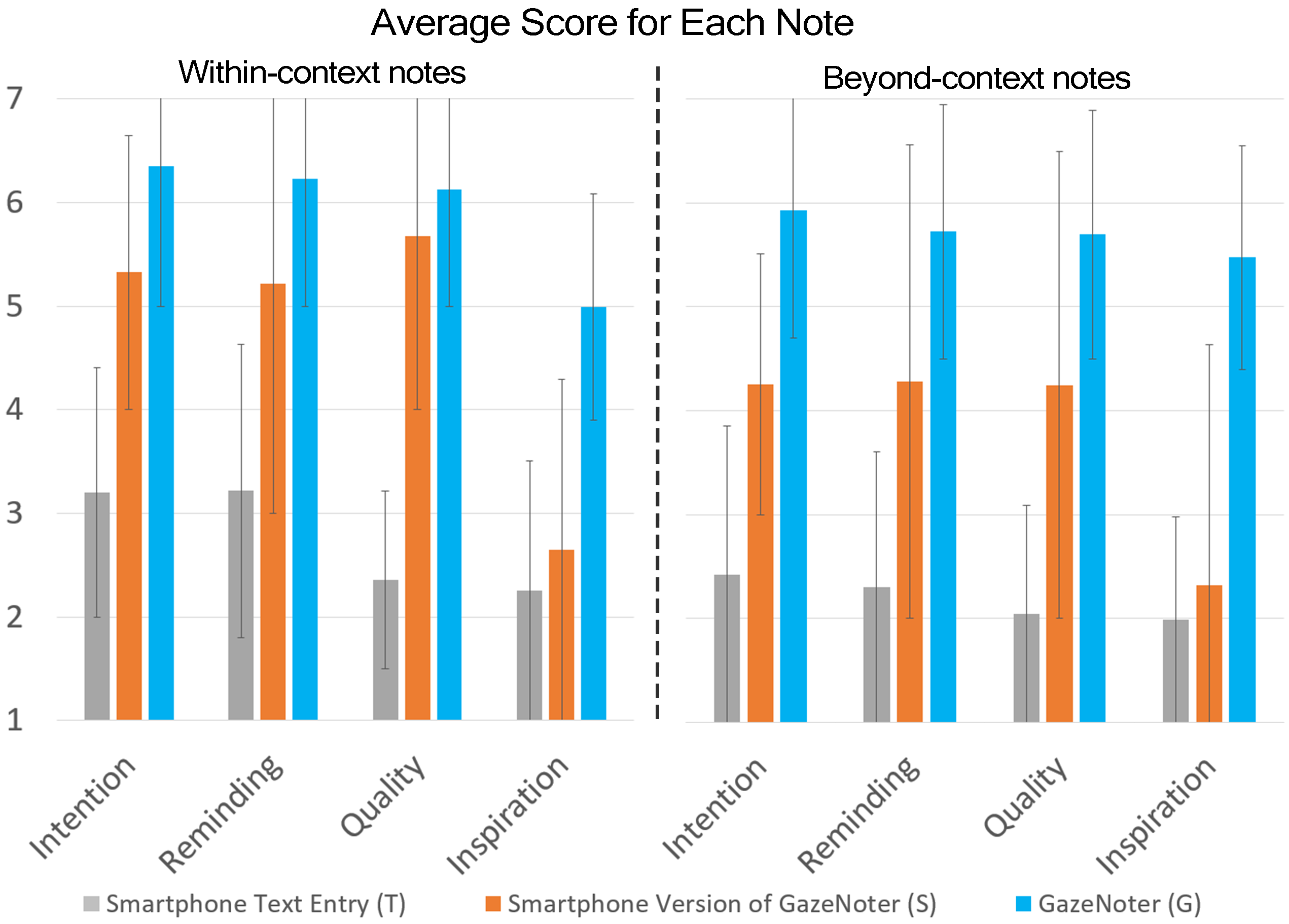}
    \caption{Score from each within-context and beyond-context note in user study 2.}
    \Description{Score from each within-context and beyond-context note in user study 2.}
    \label{fig:S2EachNote}
\end{figure}

\section{Prompts}

The prompts used in GazeNoter through the OpenAI GPT-4 API are provided. 
The text with curly braces (e.g., \textcolor{blue}{\{text\}}) in the ‘Prompt’ column is a placeholder for example input.

\begin{table*}
\centering
\begin{tabular}{l|l|l|l} 
\toprule
Prompt Goal & Prompt & Example Input(s) & \begin{tabular}[c]{@{}l@{}}
    Example\\ Response \\
 \end{tabular} \\
\hline
\begin{tabular}[c]{@{}l@{}}
Context Keyword Extraction \end{tabular} 
& 
    \begin{tabular}[c]{@{}l@{}}
    Main Sentence: \textcolor{blue}{\{New Speech Input\}}\\ 
    You are a researcher taking notes,\\
    noting down the keywords.\\
    Please extract no more than four \\
    keywords from the previous\\ 
    Main Sentence.\\ 
    Which is also shown as follows:\\
    \textcolor{blue}{\{New Speech Input\}}\\
    The extracted keywords must only \\
    exist in the Main Sentence. \\ 
    Do not extract keywords that are \\
    preposition, greeting words, or other \\
    words that are irrelevant to the sentence.\\
    Please provide me with the keywords\\
    in a format where keywords are separated\\
    by a newline, not a comma, and without\\
    an order number.\\ \end{tabular} 
& 
    \begin{tabular}[c]{@{}l@{}}
    New Speech Input: \\
    People went from city to city, \\
    holding rallies, and meetings.\\ \end{tabular} 
&
    \begin{tabular}[c]{@{}l@{}}
    people \\ 
    city \\
    rallies \\
    meetings \\ \end{tabular} 
\\
\hline
\begin{tabular}[c]{@{}l@{}}
    Derivative Keyword Derivation: \\
    Keywords exclusively related \\
    to the original keyword \end{tabular} 
&
\begin{tabular}[c]{@{}l@{}}
    Generate 2 words that are related \\
    to the word \textcolor{blue}{\{Original Keyword\}}. \\ 
    The generated words must not \\ 
    overlap with these words: \\
    \textcolor{blue}{\{Currently Displaying Context Keywords\}}.\\
    The generated words also must not \\
    have the same lemma of a word \\
    with \textcolor{blue}{\{Original Keyword\}}.\\ 
    For example, talk, talking, talked \\
    and talks all have the same lemma \\
    of a word, which is forbidden.\\ 
    Please provide me with the generated words \\
    in a format where each words are separated \\ 
    by a newline, not a comma, and \\ 
    without an order number.\\ \end{tabular}   
& 
\begin{tabular}[c]{@{}l@{}}
    Original Keyword: \\
    rallies \\ 
    \/ \\
    Currently Displaying Context Keywords: \\
    people, city, rallies, meetings 
    \\ \end{tabular}   
&   
    \begin{tabular}[c]{@{}l@{}}
    media \\ 
    civilization \\ \end{tabular} 
\\
\hline
\begin{tabular}[c]{@{}l@{}}
    Derivative Keyword Derivation: \\
    Keywords contextually \\
    associated with the \\
    original keyword \end{tabular} 
&  
\begin{tabular}[c]{@{}l@{}}
    Transcript of Speech: \textcolor{blue}{\{Previous Speech\}}. \\
    The previous paragraph is a transcript \\ 
    of a speech.\\
    Based on the above-mentioned transcript \\
    of the speech, understand the context \\ 
    of the speech and generate 2 words that \\
    are contextually related to both the \\ 
    context of the speech and also related \\
    to the word: \textcolor{blue}{\{Original Keyword\}}. \\
    The generated words must not \\ 
    overlap with these words: \\
    \textcolor{blue}{\{Currently Displaying Context Keywords\}}.\\
    The generated words also must not \\
    have the same lemma of a word \\
    with \textcolor{blue}{\{Original Keyword\}}. \\ 
    For example, talk, talking, talked \\
    and talks all have the same lemma \\
    of a word, which is forbidden.\\ 
    Please provide me with the generated words \\
    in a format where each words are separated \\ 
    by a newline, not a comma, and without an \\
    order number.\\ \end{tabular}   
    & 
    \begin{tabular}[c]{@{}l@{}}
    Original Keyword: \\
    rallies \\ 
    \/ \\
    Currently Displaying Context Keywords: \\
    people, city, rallies, meetings \\
    \/ \\
    Previous Speech: \\
    My one job was to deprogram the public \\
    from the bird lie. \\
    And deprogram is a very specific word \\ 
    because you all are programmed. \\
    We live in a pro-bird civilization \\
    drenched in propaganda. \\
    For instance, every state has a state bird, \\
    the national mascot is a bald eagle, \\
    presidents don't talk, they tweet, \\
    then the tweets are covered on the \\
    bird-logo media. \\
    Once I knew this, my first order \\
    of business was to get the information  \\
    out to the American people, \\
    get off the internet into the real world.\\ \end{tabular} 
    &  
    \begin{tabular}[c]{@{}l@{}}
    speeches \\ 
    sign \\ \end{tabular} 
    \\
\bottomrule
\end{tabular}
\end{table*}

\begin{table*}
\centering
\begin{tabular}{l|l|l|l} 
\toprule
Prompt Goal & Prompt                                               & Example Input(s) & Example Response  \\ 
\hline
\begin{tabular}[c]{@{}l@{}}
Candidate Sentence \\
Organization \end{tabular} 
& 
    \begin{tabular}[c]{@{}l@{}}
    Context: \textcolor{blue}{\{Previous Speech\}} \\
    + \\
    (If the user selected the question words in \\
    Customized Keywords) \\
    Please generate three question sentences,\\
    starting with these question words: \\
    \textcolor{blue}{\{Selected Question Words\}}. \\
    or \\
    (If the user selected ``?'' in Customized Keywords) \\
    Please generate three question sentences, \\
    but DO NOT start with these words: \\
    \textcolor{blue}{\{Question Words in Customized Keywords\}}. \\
    or \\
    (If the user did not select any Customized Keywords) \\
    Please generate three fact sentences, \\
    not question sentences. \\
    The generated sentences could be the \\ 
    brief summary of the above-mentioned \\ 
    Context paragraph. \\
    + \\
    The generated question sentences must \\
    all contain the following keywords: \\
    \textcolor{blue}{\{Selected Keywords\}}. \\
    The generated sentences must be clear \\
    and concise, not too long, no more than 10 words. \\ 
    The generated sentences must be questions \\ 
    related to the above-mentioned Context paragraph.\\ \end{tabular} 
& 
    \begin{tabular}[c]{@{}l@{}}
    Previous Speech: \\
    I went city to city, holding rallies, \\
    meeting up with our thousands of \\
    supporters, growing by the day. \\
    And I was putting up billboards \\
    wherever we went, sharing our \\
    simple but powerful \\
    message. Look how beautiful it is. \\
    Now, the government, did take note of  \\
    what we were doing and they sent \\
    some intimidators to try and deter us \\ 
    from our mission. \\
    You can see them right there. \\
    But we did not fold. \\
    We kept on going. \\
    We started holding rallies. \\
     \/ \\
    Selected Question Words: What \\
        \/ \\
    Selected Keywords: city, sign \\ \end{tabular} 
    
&
    \begin{tabular}[c]{@{}l@{}}
    What city had the \\
    most impactful signs? \\
    \/ \\
    What signs were \\
    displayed in each \\
    city? \\
    \/ \\
    What city had the \\
    most controversial \\
    signs? \\  \end{tabular}     
    \\
\bottomrule
\end{tabular}
\end{table*}
\end{document}

%% file: macro.tex
\usepackage{color}
\usepackage{algorithm}
\usepackage{algpseudocode}
\usepackage{graphicx}
\usepackage{ifthen}

\definecolor{gray}{rgb}{0.5,0.5,0.5} 
\definecolor{green}{rgb}{0, 0.4, 0} 
\definecolor{orange}{rgb}{1, 0.5, 0}     
\definecolor{mahogany}{rgb}{0.75, 0.25, 0.0}
\definecolor{purple}{rgb}{0.6, 0, 0.6}
\definecolor{purple}{rgb}{0.6, 0, 0.6}
\definecolor{darkgreen}{rgb}{0, 0.4, 0} 
\definecolor{frenchblue}{rgb}{0.0, 0.45, 0.73}

\newcommand{\ignore}[1]{}

\usepackage{color}

\newcommand{\etal}{\textit{et~al.}~}    

\newcommand{\eg}{\textit{e.g.},~}
\newcommand{\figname}{Figure}

\newboolean{revising}
\setboolean{revising}{true}
\ifthenelse{\boolean{revising}}
{

    \newcommand{\delete}[1]{}
    \newcommand{\remove}[1]{\delete{#1}}
    
    \newcommand{\robinignore}[1]{\sout{#1}}

} {
    
    \newcommand{\delete}[1]{}
    \newcommand{\remove}[1]{}

    \newcommand{\robinignore}[1]{}

}